\documentclass[prl,superscriptaddress,floatfix,showpacs,aps,reprint,letterpaper,nobalancelastpage]{revtex4-1}
\usepackage{graphicx} 
\usepackage{color} 
\usepackage{transparent}
\usepackage{amsmath}
\usepackage{hyperref} 
\usepackage{amssymb}

\newcommand{\ket}[1]{\left\lvert #1 \right\rangle}
\DeclareMathOperator{\tr}{Tr}

\newcommand{\MHz}{\mathrm{MHz}}

\newcommand{\us}{\mu\mathrm{s}}
\newcommand{\ns}{\mathrm{ns}}

\newcommand{\um}{\mu \mathrm{m}}

\begin{document}
\title{Complete universal quantum gate set approaching fault-tolerant thresholds with superconducting qubits}
\date{\today}

\author{Jerry M. Chow}
\affiliation{IBM T.J. Watson Research Center, Yorktown Heights, NY 10598, USA}
\author{Jay M. Gambetta}
\affiliation{IBM T.J. Watson Research Center, Yorktown Heights, NY 10598, USA}
\author{A. D. C\'orcoles}
\affiliation{IBM T.J. Watson Research Center, Yorktown Heights, NY 10598, USA}
\author{Seth T. Merkel}
\affiliation{IBM T.J. Watson Research Center, Yorktown Heights, NY 10598, USA}
\author{John A. Smolin}
\affiliation{IBM T.J. Watson Research Center, Yorktown Heights, NY 10598, USA}
\author{Chad Rigetti}
\affiliation{IBM T.J. Watson Research Center, Yorktown Heights, NY 10598, USA}
\author{S. Poletto}
\affiliation{IBM T.J. Watson Research Center, Yorktown Heights, NY 10598, USA}
\author{George A. Keefe}
\affiliation{IBM T.J. Watson Research Center, Yorktown Heights, NY 10598, USA}
\author{Mary B. Rothwell}
\affiliation{IBM T.J. Watson Research Center, Yorktown Heights, NY 10598, USA}
\author{J. R. Rozen}
\affiliation{IBM T.J. Watson Research Center, Yorktown Heights, NY 10598, USA}
\author{Mark B. Ketchen}
\affiliation{IBM T.J. Watson Research Center, Yorktown Heights, NY 10598, USA}
\author{M. Steffen}
\affiliation{IBM T.J. Watson Research Center, Yorktown Heights, NY 10598, USA}
\begin{abstract}
We use quantum process tomography to characterize a full universal set of all-microwave gates on two superconducting single-frequency single-junction transmon qubits. All extracted gate fidelities, including those for Clifford group generators, single-qubit $\pi/4$ and $\pi/8$ rotations, and a two-qubit controlled-NOT, exceed $95\%$ ($98\%$), without (with) accounting for state preparation and measurement errors. Furthermore, we introduce a process map representation in the Pauli basis which is visually efficient and informative. This high-fidelity gate set serves as another critical building block towards scalable architectures of superconducting qubits for error correction schemes.
\end{abstract}
\pacs{03.67.Ac, 42.50.Pq, 85.25.-j}
\maketitle

A critical prerequisite for building a scalable fault-tolerant quantum computer is the application of error correction codes \cite{Shor1996,*Steane:1996om,Bravyi1998,*Raussendorf2007,Cross2009}. In order to employ these codes, the underlying quantum gates must be performed with high fidelity above certain threshold levels. Experimentally, the onus currently lies in making scalable physical systems with universal quantum gate sets which surpass these thresholds.

Two-dimensional error-correction surface codes, with gate fidelity thresholds of $\sim$$90$-$99.5\%$ depending on measurement errors~\cite{Bravyi1998,*Raussendorf2007,Dennis2002,Cross2009}, are particularly well-suited for superconducting qubit quantum processors, as repetitive tiling of qubit and resonator networks are apposite to proposed nearest-neighbor lattices~\cite{Steffen2011}. Regarding superconducting qubits, single-qubit average gate fidelities exceeding $99\%$ have been shown~\cite{Chow2010a, Paik2011}, two-qubit gates are capable of entangled state fidelities greater than $90\%$~\cite{dicarlo_2009, Chow2011}, and three-qubit entanglement has been observed~\cite{DiCarlo2010,*Neeley2010}. Furthermore, recent experiments showing tenfold increases in coherence times of Josephson-junction qubits~\cite{Paik2011,Corcoles2011,rigetti_inprep2012} suggest that characterization of a complete high-fidelity universal set of quantum gates in superconducting qubits should be realizable.

In this Letter we report gate fidelities greater than $95\%$ for a complete universal set of gates for two fixed-frequency superconducting qubits. The qubits are single-junction transmons (SJT), coupled via a coplanar waveguide resonator~\cite{koch_charge-insensitive_2007,*schreier_suppressing_2008}. The coherence times for our two SJT device are over twice as long as those of any previously reported superconducting multi-qubit system~\cite{Chow2011}. The gates characterized in this work include the single-qubit rotations $\{I, X_{\pi},X_{\pi/2}, Y_{\pi/2}, X_{\pi/4}, X_{\pi/8}\}^{\otimes 2}$ ($R_{\theta}$ represents a rotation of angle $\theta$ around axis $R$) and a two-qubit CNOT. The combination of the $\pi/4$ rotation (commonly referred to as the $\pi/8$ gate~\cite{nielsen_chuang_2000}), the Clifford group generators $\{X_{\pi/2},Y_{\pi/2}\}$, and the two-qubit CNOT form a universal set~\cite{Shor1996}. The two-qubit CNOT between the fixed-frequency qubits is implemented using the cross-resonance (CR) interaction \cite{Paraoanu2006,*Rigetti2010,Chow2011,Groot2012}. Although the gate fidelities are obtained via standard quantum process tomography (QPT) of each respective gate, in this work we introduce a different, but efficient visual representation of process maps, the Pauli transfer matrix, $\mathcal{R}$. $\mathcal{R}$ describes the action of a process on the components of the density matrix represented in the basis of Pauli operators
and helps establish a number of properties of the underlying process which are otherwise hidden in the standard chi-matrix representation~\cite{nielsen_chuang_2000}. We find slight variations in the fidelities across the different gates and, from the associated maps, identify that our errors are not coherence limited but rather control limited and inherent in the QPT scheme. This suggests that as superconducting qubits continue to improve in coherence times, the adoption of longer control sequences inspired by nuclear magnetic resonance quantum computing \cite{Cummins2003}, and other efficient gate characterization schemes, will become necessary.

\begin{figure}[htbp!]
\centering
\includegraphics[width=0.47\textwidth]{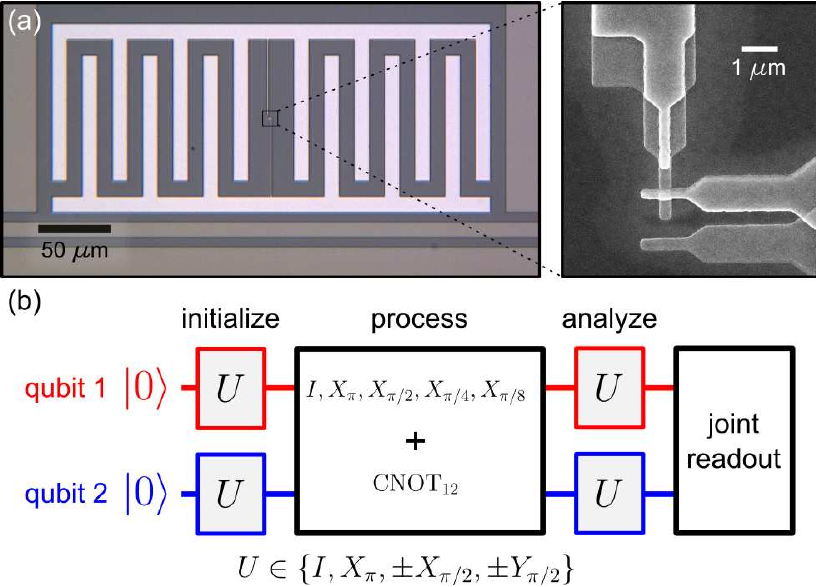}
\caption{\label{fig:1} (color online)~Single-junction transmon and quantum process tomography pulse sequence.~(a) Optical micrograph of single-junction transmon device. Design is similar to standard transmon devices, with interdigitated capacitors shunting the Al/AlOx Josephson junction on either side. Inset: scanning-electron micrograph of Josephson junction, with critical current $I_0 = 28$ nA. (b) Pulse sequence for performing quantum process tomography. Over-complete set of rotations $\{I, X_{\pi}, X_{\pm\pi/2}, Y_{\pm\pi/2}\}$ are used to generate input states and to analyze the process before joint readout via the cavity.}
\end{figure}

The transmon qubit has become a popular choice for superconducting quantum computing applications due to its excellent coherence properties \cite{koch_charge-insensitive_2007,*schreier_suppressing_2008}. Implementations of transmons in planar qubit-cavity devices have traditionally been formed via capacitively shunting a Cooper-pair box, preserving tunability of the qubit transition frequencies. The ability to dynamically tune qubit frequencies is necessary for a number of entangling gate schemes~\cite{dicarlo_2009,Ansmann:2009yq} but often involves fast flux-biasing to frequencies with reduced coherence. In this work, we focus only on fixed-frequency SJTs, having previously generated entangled states using an all-microwaves scheme \cite{Chow2011}.

Two SJTs [Fig.~\ref{fig:1}(a)] with transition frequencies $\omega_1/2\pi = 5.0554$ GHz and $\omega_2/2\pi = 4.9895$ GHz are coupled via a 7.325 GHz coplanar waveguide resonator. Single-qubit rotations are performed by irradiating each qubit through independent on-chip microwave bias lines at the respective transition frequency. The relaxation times of the two qubits are measured to be $T_1^{(1)} = 8.2~\us$ and $T_1^{(2)} = 9.7~\us$ with Ramsey-fringe coherence times of $T_2^{(1)}= 7.1$ and $T_2^{(2)} = 10.3~\us$. We attribute the improved coherence times to reduced surface loss contributions~\cite{Wenner2011a} via larger qubit feature sizes [$10~\um$ capacitive shunt fingers and gaps, see Fig.~\ref{fig:1}(a)] and meticulous radiation shielding techniques~\cite{Barends2011,Corcoles2011}. Further sample details are given in supplementary material~\cite{chow_supp_univ}.

We employ a series of repeated pulse experiments to accurately calibrate microwave amplitudes, offsets, and phases of single-qubit rotations~\cite{chow_supp_univ}. The single-qubit rotations are shaped gaussian envelopes (gaussian standard deviation $\sigma = 10~\ns$, total gate length $4\sigma$) with derivative gaussian quadrature corrections to account for excited state leakage \cite{Motzoi:2009ca,Chow2010a}. 

The two-qubit interaction is the CR effect and its implementation and advantages as a multi-qubit gate scheme are previously detailed in Refs.~\onlinecite{Rigetti2010,Chow2011}. To review, the $\text{CR}_{ij}$ effect is exhibited as a qubit $i$ (control) state dependent drive of qubit $j$ (target). In our system, we drive microwaves resonant with qubit 2 onto the microwave bias line addressing qubit 1. A residual classical cross-talk term is present, characterized by the parameter $m_{12} = 0.22$, which is the fraction of the direct Rabi frequency  experienced by qubit 2 through driving on qubit 1. The strength of the desired two-qubit quantum effect is bounded by the residual qubit-qubit cavity-mediated dispersive interaction, which we determine to be $J/\pi = 7.44\,\MHz$ from detailed spectroscopy of transmon energy transitions and comparison to diagonalization of the Hamiltonian for our multi-level system. The experimental tune-up and implementation of a CNOT gate with this interaction follows that of Ref.~\onlinecite{Chow2011} and in this work the pulse-shape is a flat-top gaussian ($\sigma = 10~\ns$) of length 110 ns. 

\begin{figure}[tbp!]
\centering
\includegraphics[width=0.47\textwidth]{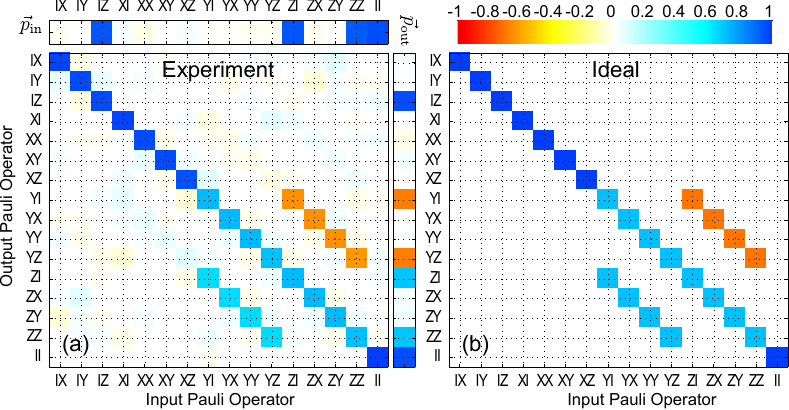}
\caption{\label{fig:2}~Quantum process tomography for $X_{\pi/4} \otimes I$ represented as the Pauli transfer matrix $\mathcal{R}$. (a)~Experimentally extracted $\mathcal{R}$ for $X_{\pi/4} \otimes I$ with gate fidelity $F_{\text{g}} = 0.9687$. To illustrate the action of $\mathcal{R}$, an input state $\ket{00}$ (state fidelity $F_{\text{s}}=0.9818$) is shown above $\mathcal{R}$, and the output state $\cos(\pi/8)\ket{00}-i\sin(\pi/8)\ket{10}$ ($F_{\text{s}}= 0.9969$) is shown to the right. (b)~Ideal $\mathcal{R}$ for $X_{\pi/4} \otimes I$.}.
\end{figure}

To measure the two-qubit state, we employ a series of joint two-qubit measurements via non-linear driving of the cavity \cite{Reed2010}. Calibrating this joint measurement~\cite{chow_supp_univ}, we perform state tomography on arbitrary two-qubit states through the application of an overcomplete set of analysis pulses $\{I, X_{\pi}, X_{\pm\pi/2},  Y_{\pm\pi/2}\}^{\otimes 2}$. An efficient maximum-likelihood estimation (MLE) algorithm \cite{Smolin2012} is used to compute two-qubit states which can be represented by the Pauli state vector $\vec{p}$~\cite{Chow2010}. The elements of $\vec{p}$ are the expectation values of the two qubit Pauli operators, $\langle UV\rangle$, where $U,V \in {I,X,Y,Z}$.

QPT is accomplished through the compilation of an overcomplete set of 36$\times$36 measurements [Fig.~\ref{fig:1}(b)]. State tomography as described above, is performed for the 36 different input states generated by $\{I, X_{\pi}, X_{\pm\pi/2}, Y_{\pm\pi/2}\}^{\otimes 2}$. Instead of  the standard chi-matrix representation of the process map~\cite{nielsen_chuang_2000}, we present the Pauli transfer matrix $\mathcal{R}$, which maps an input Pauli state vector $\vec{p}_{\text{in}}$ to an output Pauli state vector $\vec{p}_{\text{out}}$, $\vec{p}_{\text{out}} = \mathcal{R}\vec{p}_{\text{in}}$. $\mathcal{R}$ is obtained through a semi-definite program taking into account the covariance matrix for the different independent measurements. Here, the semi-definite program extraction of $\mathcal{R}$ weights the measurements unequally~\cite{chow_supp_univ}. The gate fidelity $F_{\text{g}}$ is calculated from the $\mathcal{R}$ map by $F_{\text{g}} = (\tr[\mathcal{R}^{\dagger}\mathcal{R}]+d)/(d^2+d)$ with $d=2n$ and $n$ is the number of qubits. 

\begin{figure}[tbp!]
\centering
\includegraphics[width=0.47\textwidth]{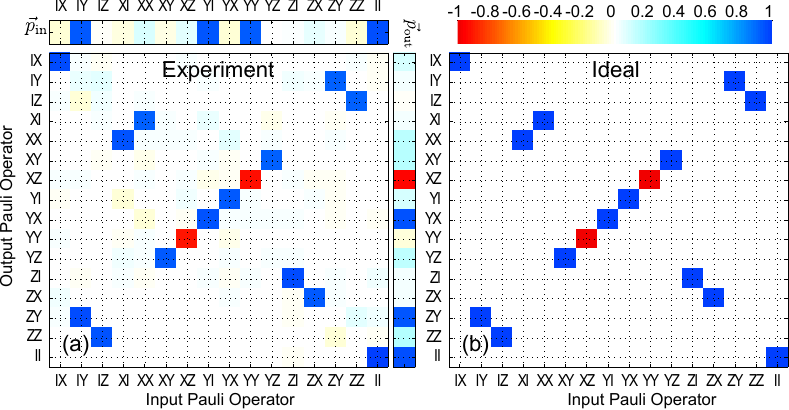}
\caption{\label{fig:3}~Quantum process tomography for $\text{CNOT}_{12}$ represented as the Pauli transfer matrix $\mathcal{R}$. (a)~Experimentally extracted $\mathcal{R}$ for $\text{CNOT}_{12} $ with gate fidelity $F_{\text{g}} = 0.9507$. To illustrate the action of $\mathcal{R}$, an input state $(\ket{0}+i\ket{1})(\ket{0}+i\ket{1})/2$ ($F_{\text{s}}= 0.9787$) is shown above $\mathcal{R}$, and the output entangled state $(\ket{00}+i\ket{01}-\ket{10}+i\ket{11})/2$ ($F_{\text{s}} = 0.9827$, $\mathcal{C} = 0.994$) is shown to the right. (b)~Ideal $\mathcal{R}$ for $\text{CNOT}_{12}$.}
\end{figure}

We perform QPT for the processes corresponding to the identity operation $I^{\otimes 2}$, all independent single-qubit rotations of $\pi$, $\pi/2$ (around $x$- and $y$-axes), $\pi/4$, and $\pi/8$, as well as the $\text{CNOT}_{12}$ operation. The experimentally extracted map $\mathcal{R}$ for the universal single-qubit $\pi/4$ rotation for qubit 1 is shown in  Fig.~\ref{fig:2}(a) and the ideal $\mathcal{R}_{\text{ideal}}$ is shown in Fig.~\ref{fig:2}(b). From $\mathcal{R}$, we estimate $F_{\text{g}} = 0.9687$. The Pauli-transfer matrix $\mathcal{R}$ for the two-qubit entangling $\text{CNOT}_{12}$ gate is shown in Fig.~\ref{fig:3}(a), from which we estimate $F_{\text{g}} = 0.9507$. Within this set the highest entangled state fidelity ($F_{\text{s}} = 0.9827$) corresponds to the state $(\ket{00}+i\ket{01}-\ket{10}+i\ket{11})/2$, with an associated concurrence $\mathcal{C} = 0.994$. The extracted $F_{\text{g}}$ for the complete set of gates are given in Table~\ref{table:1}.

Figures~\ref{fig:2}(a) and~\ref{fig:3}(a) also demonstrate the action of the different Pauli transfer maps on specific input states. Above the experimentally extracted $\mathcal{R}$ are the measured Pauli state-vectors corresponding to $\ket{00}$ and $(\ket{0}+i\ket{1})(\ket{0}+i\ket{1})/2$ in Fig.~\ref{fig:2}(a) and Fig.~\ref{fig:3}(a), respectively. We can follow all the elements of each of these Pauli states downwards in the figures into $\mathcal{R}$, which transfers these weights over into the Pauli operators to the right of $\mathcal{R}$. Therefore, the Pauli state to the right reflects the output state given the operation of the map $\mathcal{R}$ on the input Pauli state vector, and this visualization very simply demonstrates its effect.

\begin{table}
\begin{ruledtabular}
\begin{tabular}{|l|c|c|c|c|}
Gate & $F_{\text{g}}$ & $\Delta F_{\text{g}}~(\times 10^{-4}$) & $F_{\text{pure}}$ & $\Upsilon_{\text{np}}$ \\ 
\hline
$I\otimes I$ 			& 0.9691 	& 3.6 	& 0.9954	& 0.036\\
$X_{\pi} \otimes I$ 	& 0.9618	& 4.3 	& 0.9935	& 0.029\\ 
$X_{\pi/2} \otimes I$	& 0.9620	& 5.2	& 0.9955	& 0.030\\
$Y_{\pi/2} \otimes I$	& 0.9621	& 5.3	& 0.9956	& 0.046\\
$X_{\pi/4} \otimes I$	& 0.9687	& 5.5	& 0.9962	& 0.038\\
$X_{\pi/8} \otimes I$	& 0.9649	& 5.2 	& 0.9962	& 0.038\\
$I\otimes X_{\pi} $ 	& 0.9629	& 4.4	& 0.9906	& 0.033\\ 
$I\otimes X_{\pi/2}$	& 0.9597	& 3.9	& 0.9955	& 0.031\\
$I\otimes Y_{\pi/2} $	& 0.9569	& 4.6	& 0.9961	& 0.040\\
$I\otimes X_{\pi/4}$	& 0.9644	& 4.6	& 0.9963	& 0.035\\
$I\otimes X_{\pi/8}$	& 0.9666	& 6.0 	& 0.9968	& 0.042\\
$\text{CNOT}_{12}$		& 0.9507	& 6.5	& 0.9968	& 0.035 
\end{tabular}
\end{ruledtabular}
\caption{\label{table:1}Summary of gate fidelity $F_{\text{g}}$, statistical error $\Delta F_{\text{g}}$, purified fidelity $F_{\text{pure}}$, and non-physical error $\Upsilon_{\text{np}} = 0.5\|\mathcal{R}_\mathrm{MLE}-\mathcal{R}_\mathrm{exp}\|_2$ for complete universal set of gates on two qubits.}
\end{table}

The $\mathcal{R}$ representation is a more efficient and intuitive method for representing a quantum operation than the standard chi-matrix \cite{nielsen_chuang_2000}, as it consists of only real numbers and possesses a few other nice visual properties. First, it is simple to tell if the map is trace preserving, which amounts to $\mathcal{R}_{II,jk} = \delta_{Ij}\delta_{Ik} $ for all $j,k\in\{I,X,Y,Z\}$. Next, we can also determine if the map is unital, if $\mathcal{R}_{jk,II} =  \delta_{Ij}\delta_{Ik} $ for all $j,k\in \{I,X,Y,Z\}$. Finally, the elements in $\mathcal{R}$ are bounded by $\pm1$ and for any Clifford operation there is exactly one non-zero element in each row and column with unit magnitude. 

It is possible to further investigate the $\mathcal{R}$ maps to understand the errors in our system and determine whether the loss in fidelity for all of the gates is due to statistical or systematic errors. For statistical errors, we use a bootstrapping method which generates new realizations of tomography experiments based off the variance of our measurement operator calibrations~\cite{chow_supp_univ}. From these generated experiments, new ensembles of maximum-likelihood estimates are obtained and the variance of these ensembles serve as an upper bound on statistical fluctuations of our estimated gate errors~\cite{chow_supp_univ}. For all of the gates studied in this work, the statistical component is found to be $\sim 3-6\times 10^{-4}$ (Table~\ref{table:1}), much smaller than the QPT extracted gate errors of $\approx 5\%$. 

The small value for statistical fluctuations suggests that our primary sources of error are systematic in nature. These can include decoherence, over- or under-rotations, and phase errors, all of which can occur during the actual processes to be characterized, or during the state preparation and measurement analysis gates. Given the coherence and gate times of the process tomography sequence, we estimate a total error of $1.62\%$ for single qubit gates and $2.55\%$ for the CNOT, which is still smaller than the observed error and not the primary source. 

The remaining error can be accounted for via miscalibration of $1\%$ in the single-qubit gates used for preparation and analysis. Simulations of the $\mathcal{R}$ maps including this level of calibration error agree well with the non-ideal elements found in the $\mathcal{R}$ data ~\cite{chow_supp_univ}. This is further confirmed through the calculation of the purified fidelity, $F_{\text{pure}}$, which estimates how close the unitary contribution of the map is to the ideal gate.  $F_{\text{pure}}$ is defined as the overlap of the ideal map with the maximum eigenvector of the Choi matrix \cite{Choi1975} corresponding to ${\mathcal{R}}_{\rm MLE}$ \cite{chow_supp_univ}. For all of the gates investigated, $F_{\text{pure}} > 0.99$ (Table~\ref{table:1}) indicating that we perform the correct unitary to better than $\approx1\%$, but we are adding on non-purity conserving operations due to systematic errors. These errors can be quantified via $\Upsilon_{\text{np}} = 0.5\|\mathcal{R}_{\mathrm{MLE}}-\mathcal{R}_{\mathrm{exp}}\|_2$, which is half the 2-norm distance between the raw experimentally determined $\mathcal{R}_{\text{exp}}$  and the most likely physical map $\mathcal{R}_{\text{MLE}}$. For all gates $\Upsilon_{\text{np}} \approx 3-4\%$ (Table \ref{table:1}), which suggests that propagation of small errors in our preparation and measurement operations in the QPT accounts for a large portion of the gate errors obtained experimentally \cite{chow_supp_univ}.

\begin{table}
    \begin{ruledtabular}
    \begin{tabular}{lc}
   Mechanism & Error    \\
    \hline
    Statistical errors ($\Delta F_{\text{g}}$) & $0.1\%$ \\ 
   	$T_1/T_2$ (simulation from $1-F_{\text{g}}$) & $1.5-2.6\%$ \\
	Unitary error in gate (1-$F_\mathrm{pure}$) & 0.5-1$\%$\\
	Non physical errors ( $\Upsilon_{\text{np}}$) &3$\%$\\
	Total (1-$F_{\text{g}}$) & 3.5-5$\%$ 
 \end{tabular}
    \end{ruledtabular}
    \caption{\label{table:2}~The error budget. }
\end{table}

Table~\ref{table:2} summarizes the different mechanisms to which we attribute our gate error budget. Further discussion regarding these error metrics are given in the supplementary material \cite{chow_supp_univ}. Our QPT results further reinforce the importance of other methods for gate characterization, such as randomized benchmarking~\cite{Magesan2011}, as $F_{\text{g}}$ can be significantly misrepresented due to small systematic errors in preparation and analysis. 

\begin{figure}[tbp!]
\centering
\includegraphics[width=0.45\textwidth]{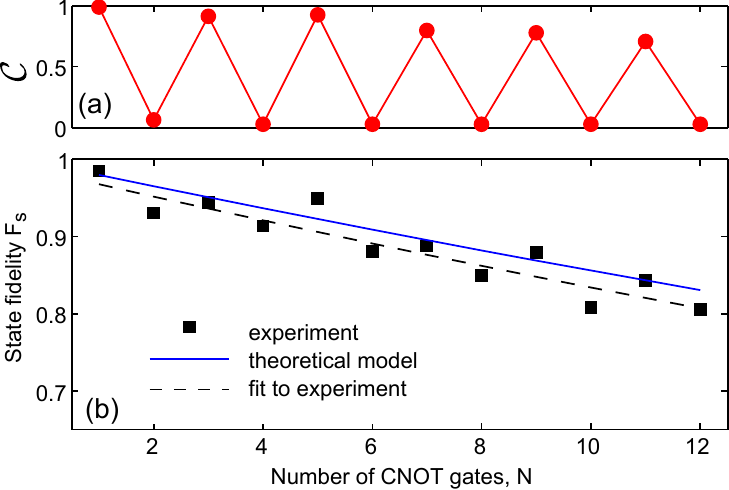}
\caption{\label{fig:4} (color online)~Concurrence and state fidelity after $N$ applications of $\text{CNOT}$. (a)~Concurrence $\mathcal{C}$ of final states after applying $N$ CNOT gates to input state $(\ket{00}+i\ket{10})/\sqrt{2}$. (b)~Black squares are final state fidelity $F_\text{s}$ of experimentally obtained states to ideal Bell (original input) state for odd (even) $N$. Dashed black line is a fit to the data assuming a model of $AF_{\text{g}}^N+B$, where $A$, $B$, and gate fidelity $F_{\text{g}}$ are fit parameters. The fit gives an error per gate $1-F_{\text{g}} = 0.0164$. Solid blue line is a theoretical model using the measured coherence values and gate times.}
\end{figure}

Finally, as a precursor to full randomized benchmarking sequences, we show $\mathcal{C}$ and $F_{\text{s}}$, after the repeated application of up to 12 CNOT gates on the input state generated by $X_{-\pi/2}\otimes I$.  Figure~\ref{fig:4}(a) reveals $\mathcal{C}$ for up to 12 applications of CNOT, indicating as desired no entanglement for even numbers ($N$) of gates. In Fig.~\ref{fig:4}(b), we plot the $F_{\text{s}}$ of the final state to the ideal states, the Bell state $(\ket{00}+i\ket{11})/\sqrt{2}$ for odd $N$ and the input state $(\ket{00}+i\ket{10})/\sqrt{2}$ for even $N$. By assuming an exponential model for the state fidelity, $F_{\text{s}} = AF_{\text{g}}^N+B$, with $A$ and $B$ as fit parameters (dashed black line), we extract a gate fidelity $F_{\text{g}} = 0.9836$.  We find that this is in good agreement with a simulation of $F_{\text{g}}$ taking into account the coherence parameters of the system and the durations of the gates involved in the experiment (solid blue line). 

Thus, we have shown a complete universal set of high-fidelity gates on two high-coherence fixed-frequency superconducting qubits. Although the gate fidelities obtained via QPT are at the $95\%$ level, both the analysis of the $\mathcal{R}$ maps with only $1\%$ systematic calibration errors and the repeated CNOT sequences suggest that the intrinsic gate fidelities are $>98\%$, near the coherence time limit. Moving forward, finer pulse calibration tune-up sequences must be implemented and gate errors will be characterized via randomized benchmarking to avoid systematic errors in state preparation and analysis. Nonetheless, this demonstration further paves the road towards $>10$ qubits for implementing more complicated pieces of quantum error correction.

\begin{acknowledgments}
We acknowledge experimental contributions from J. Rohrs, B. R. Johnson, T. Ohki, J. Strand, and B.L.T. Plourde. We acknowledge support from IARPA under contract W911NF-10-1-0324. All statements of fact, opinion or conclusions contained herein are those of the authors and should not be construed as representing the official views or policies of the U.S. Government.
\end{acknowledgments}

\end{document}


\title{Supplementary material for `Complete universal quantum gate set approaching fault-tolerant thresholds with superconducting qubits'}
\date{\today}

\author{Jerry M. Chow}
\affiliation{IBM T.J. Watson Research Center, Yorktown Heights, NY 10598, USA}
\author{Jay M. Gambetta}
\affiliation{IBM T.J. Watson Research Center, Yorktown Heights, NY 10598, USA}
\author{A. D. C\'orcoles}
\affiliation{IBM T.J. Watson Research Center, Yorktown Heights, NY 10598, USA}
\author{Seth T. Merkel}
\affiliation{IBM T.J. Watson Research Center, Yorktown Heights, NY 10598, USA}
\author{John A. Smolin}
\affiliation{IBM T.J. Watson Research Center, Yorktown Heights, NY 10598, USA}
\author{Chad Rigetti}
\affiliation{IBM T.J. Watson Research Center, Yorktown Heights, NY 10598, USA}
\author{S. Poletto}
\affiliation{IBM T.J. Watson Research Center, Yorktown Heights, NY 10598, USA}
\author{George A. Keefe}
\affiliation{IBM T.J. Watson Research Center, Yorktown Heights, NY 10598, USA}
\author{Mary B. Rothwell}
\affiliation{IBM T.J. Watson Research Center, Yorktown Heights, NY 10598, USA}
\author{J. R. Rozen}
\affiliation{IBM T.J. Watson Research Center, Yorktown Heights, NY 10598, USA}
\author{Mark B. Ketchen}
\affiliation{IBM T.J. Watson Research Center, Yorktown Heights, NY 10598, USA}
\author{M. Steffen}
\affiliation{IBM T.J. Watson Research Center, Yorktown Heights, NY 10598, USA}
\date{\today}
\maketitle

\section{Sample parameters and calibration}

The single-junction transmon ground to first excited-state frequencies are at $f_1^{01} = 5.0554$ GHz and $f_2^{01} = 4.9895$ GHz. From spectroscopy of the system with varying powers, we locate multi-photon transitions up to the third excited-state: $f_1^{02}/2 = 4.9454$ GHz, $f_1^{03}/3 = 4.8296$ GHz,  $f_2^{02}/2 = 4.8753$ GHz, $f_1^{03}/3 = 4.7497$ GHz. We estimate the crosstalk parameter $m_{12}=0.22$ from Rabi driving qubit 2 via the bias line associated with qubit 1 and comparing with the same drive but applied directly to qubit 2. From all the transitions of the two transmons, by using a model assuming transverse coupling between the two qubits, we estimate the coupling strength to be $J/2\pi = 7.44$ MHz. We also estimate a residual $ZZ$  interaction of $1.1$ MHz between the qubits via Ramsey fringe experiments with and without one of the qubits in the excited state.

Single-qubit $\pi/2$ and $\pi$ pulses are calibrated using sets of repeated sequences. First, the $\pi/2$ amplitudes are tuned up by measuring the cavity response after $\{X_{\pi/2},[X_{\pi/2},X_{\pi/2}]^{N}\}$ for $N$ up to 30. Then, we bootstrap off the tuned up $\pi/2$ pulse to find the appropriate $\pi$ amplitude via the pulse train $\{X_{\pi/2},[X_{\pi}]^N\}$, again for $N$ up to 30 pulses. Finally, the derivative-pulse shaping (DRAG) parameters~\cite{Chow2010a} and gain imbalance between $x$ and $y$ rotations are tuned via the experiments, $\{X_{\pi/2},[X_{\pi/2}, -X_{\pi/2}]^N\}$ and $\{X_{\pi/2},[X_{\pi/2},Y_{\pi/2},-Y_{\pi/2},-X_{\pi/2}]^N\}$ also for $N$ up to 30.

\section{Representations of a map}

Defining $\mathcal{H}_d$ as a Hilbert space of dimensions $d$, and $\mathcal{L}(\mathcal{H}_d)$ as the space of linear operators on $\mathcal{H}_d$, then the Choi-Jamiolkowski isomorphism states that any 
map $\Lambda: \mathcal{L}(\mathcal{H}_d)\rightarrow \mathcal{L}(\mathcal{H}_d)$ can be represented by the Choi matrix  
\begin{equation}\rho_\Lambda =\frac{1}{d} \sum_{i,j} E_{ij} \otimes\Lambda(E_{ij}).\end{equation} Here $\rho_\Lambda \in\mathcal{L}(\mathcal{H}_d\otimes\mathcal{H}_d)$ and $E_{ij}$ is the matrix with 1 in the $ij^\mathrm{th}$
 entry and 0's elsewhere. In ref.~\onlinecite{Choi1975}, Choi proved that $\Lambda$ is completely positive if and only if $\rho_\Lambda$ is positive semidefinite.  

The Choi matrix allows us to test if the map is trace preserving by 
$\mathrm{Tr}_\mathrm{B}[\rho_\Lambda]=I/d$ and if the map is unital by $\mathrm{Tr}_\mathrm{A}[\rho_\Lambda]=I/d$. Furthermore, the action of the map on any state $\rho\in\mathcal{L}(\mathcal{H}_d)$ is given by
\begin{equation}\Lambda(\rho) = d \mathrm{Tr}_\mathrm{A}[(\rho^T\otimes I)\rho_\Lambda]. \label{mapopt}\end{equation}

An alternative representation of a map is the Pauli transfer matrix, $\mathcal{R} $, defined by
\begin{equation}
\mathcal{R}_{ij} = \frac{1}{d} {\rm Tr} (P_i \Lambda(P_j)).
\end{equation}
This representation is simply the Choi matrix expanded in the Pauli operator basis (here the $P_j$'s represent some ordering of the strings of Pauli operators $\{I,\sigma_x,\sigma_y,\sigma_z \} $), since it is easy to show that
\begin{equation}
\mathcal{R}_{ij} = \mathrm{Tr}[\rho_\Lambda P_j^T\otimes P_i], 
\end{equation}
and therefore 
\begin{equation}
\rho_\Lambda = \frac{1}{d^2}\sum_{i,j} \mathcal{R}_{ij} P_j^T\otimes P_i. 
\end{equation}
This is a useful representation as the act of the map on a state $\rho$ in the Pauli representation ($\rho = \frac{1}{d}  \vec{p}\cdot \vec{P}$) is a simple matrix transformation of the Pauli state-vector $ \vec{p}$ given by
\begin{equation}
 \vec{p}' =  \mathcal{R} \vec{p},
\end{equation}
which is obvious from our definition of $\mathcal{R}$.  Therefore, the Pauli transfer matrix is also related to  the Liouville or superoperator representation by a simple basis transformation (the transformation from $E_{ij} $ to $P_k$) and composition of maps is described simply by matrix multiplication.   

The $\mathcal{R}$ matrix has a few nice properties when compared to the standard chi-matrix \cite{Chuang1997} for determining whether a map is physical.  First, $\mathcal{R}$ consists of only real numbers and so contains exactly the same number of parameters as the gate, with no redundancy due to Hermiticity.  Second, it is easy to see what operators are mapped to and from the identity by looking at the final row and column respectively (assuming $P_{ d^2} = I$).  If only the identity is mapped back onto itself, then the map is trace-preserving, and if the identity remains the identity the map is unitial.  This corresponds to a final row or column of all zeros and a single one. Lastly, the elements of  $\mathcal{R}$ are bounded by $\pm1$, and for any Clifford operations there is exactly one non-zero element in each row and column with magnitude one, as Clifford operators simply map Pauli operators to other Pauli operators. 

\begin{figure*}
\centering
\begin{tabular}{|c|c|}\hline\\
\includegraphics[width=0.4\textwidth]{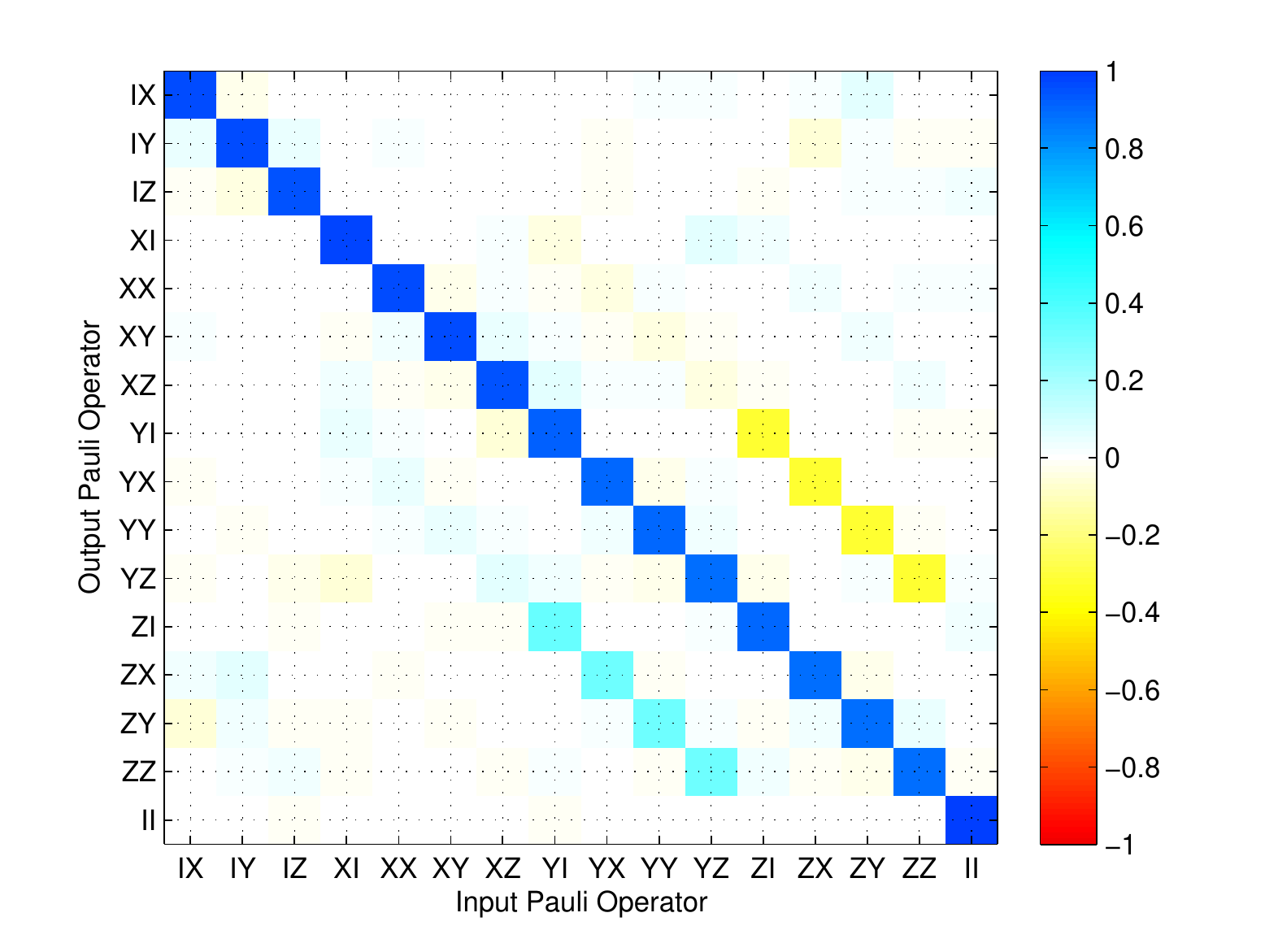} &\includegraphics[width=0.4\textwidth]{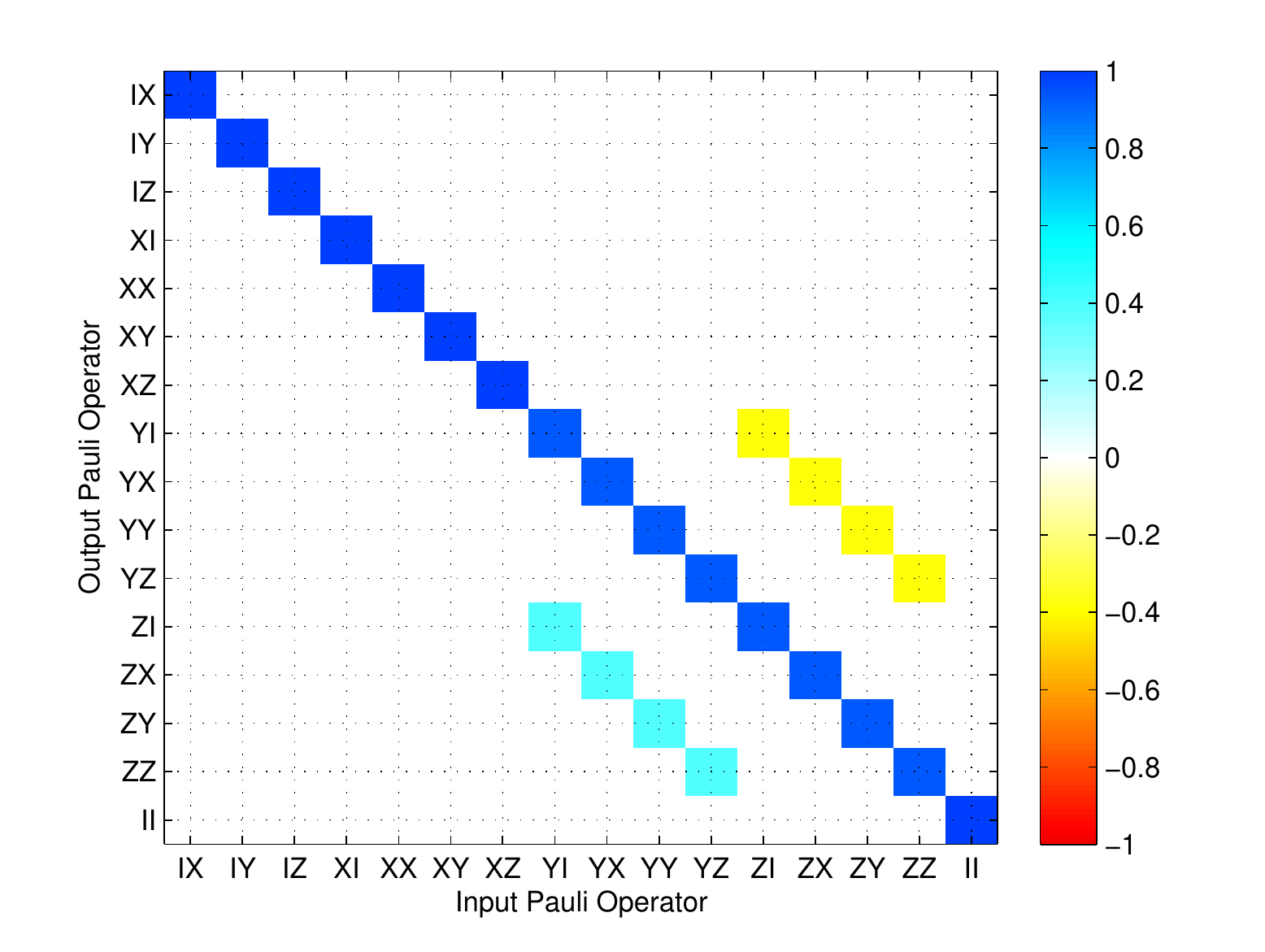}\\
\includegraphics[width=0.4\textwidth]{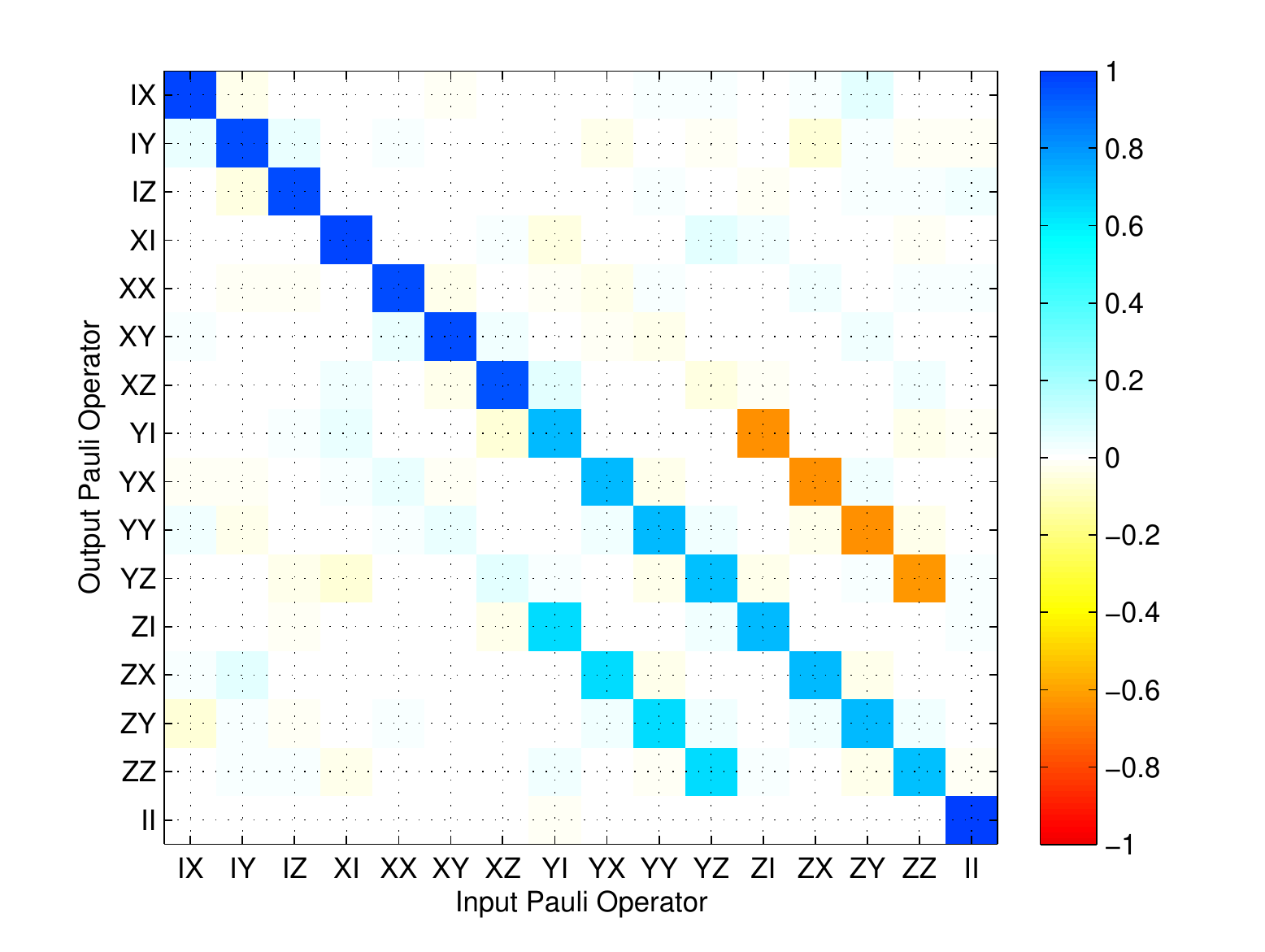} &\includegraphics[width=0.4\textwidth]{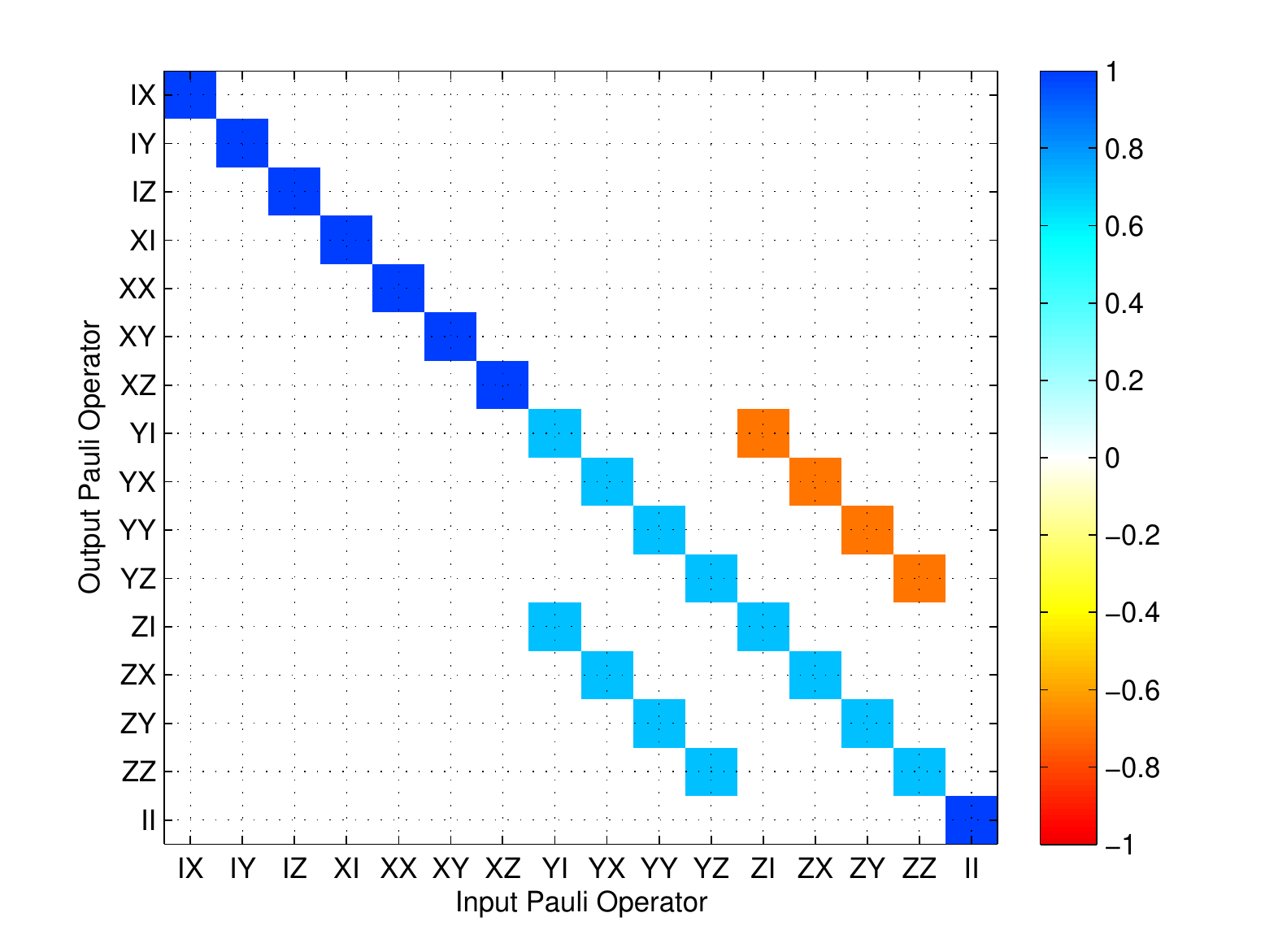}\\
\includegraphics[width=0.4\textwidth]{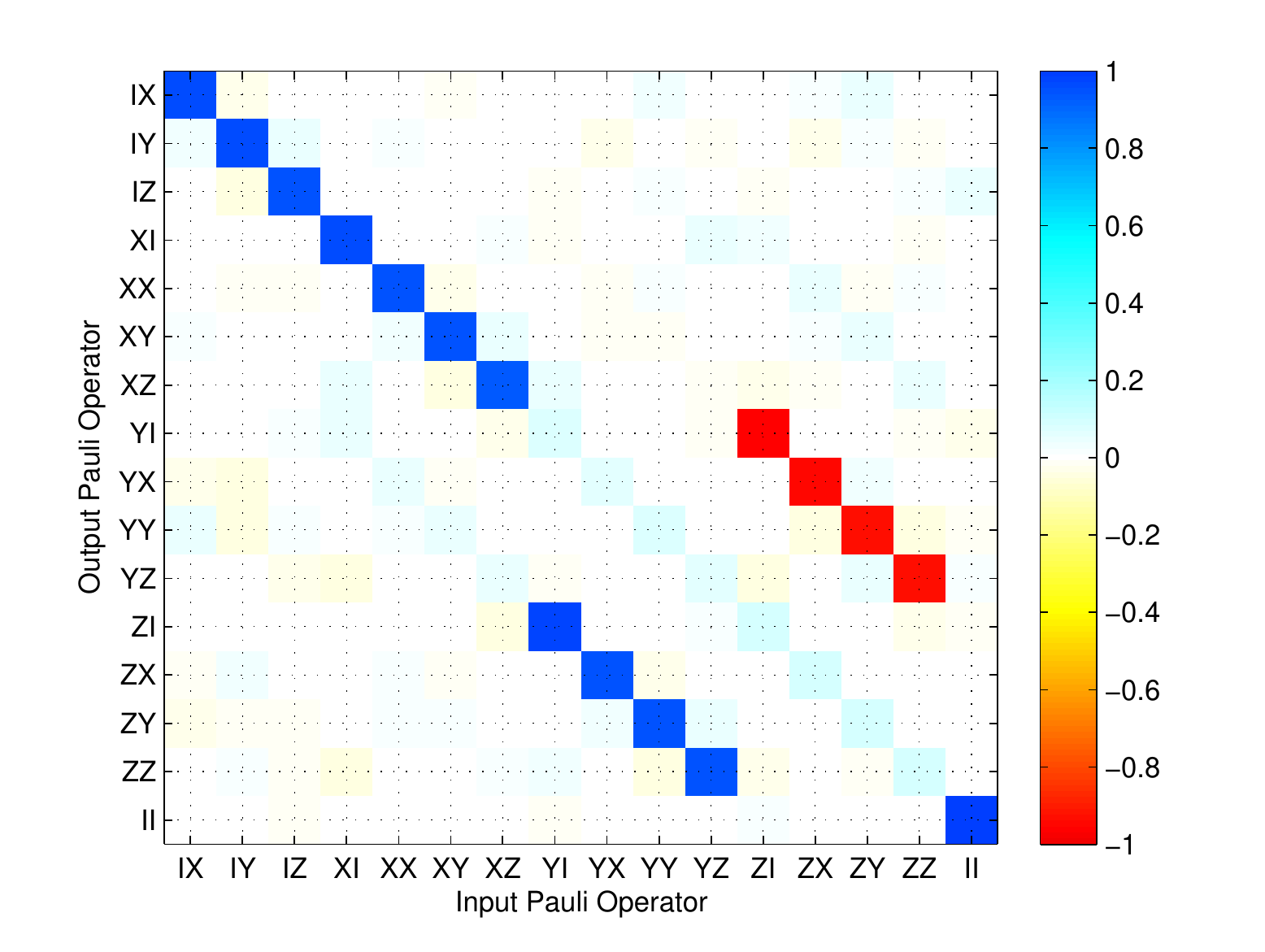} &\includegraphics[width=0.4\textwidth]{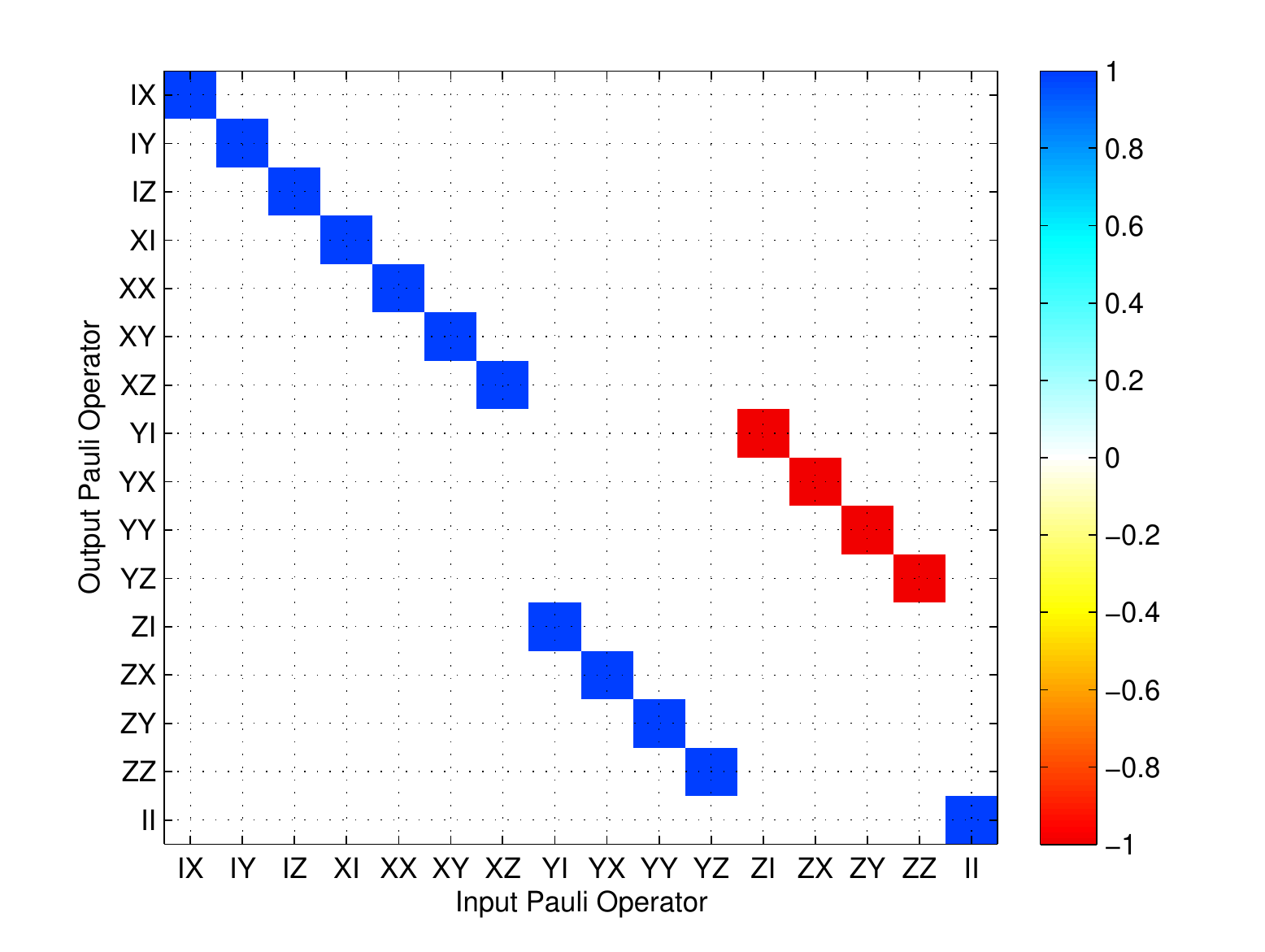}\\
\includegraphics[width=0.4\textwidth]{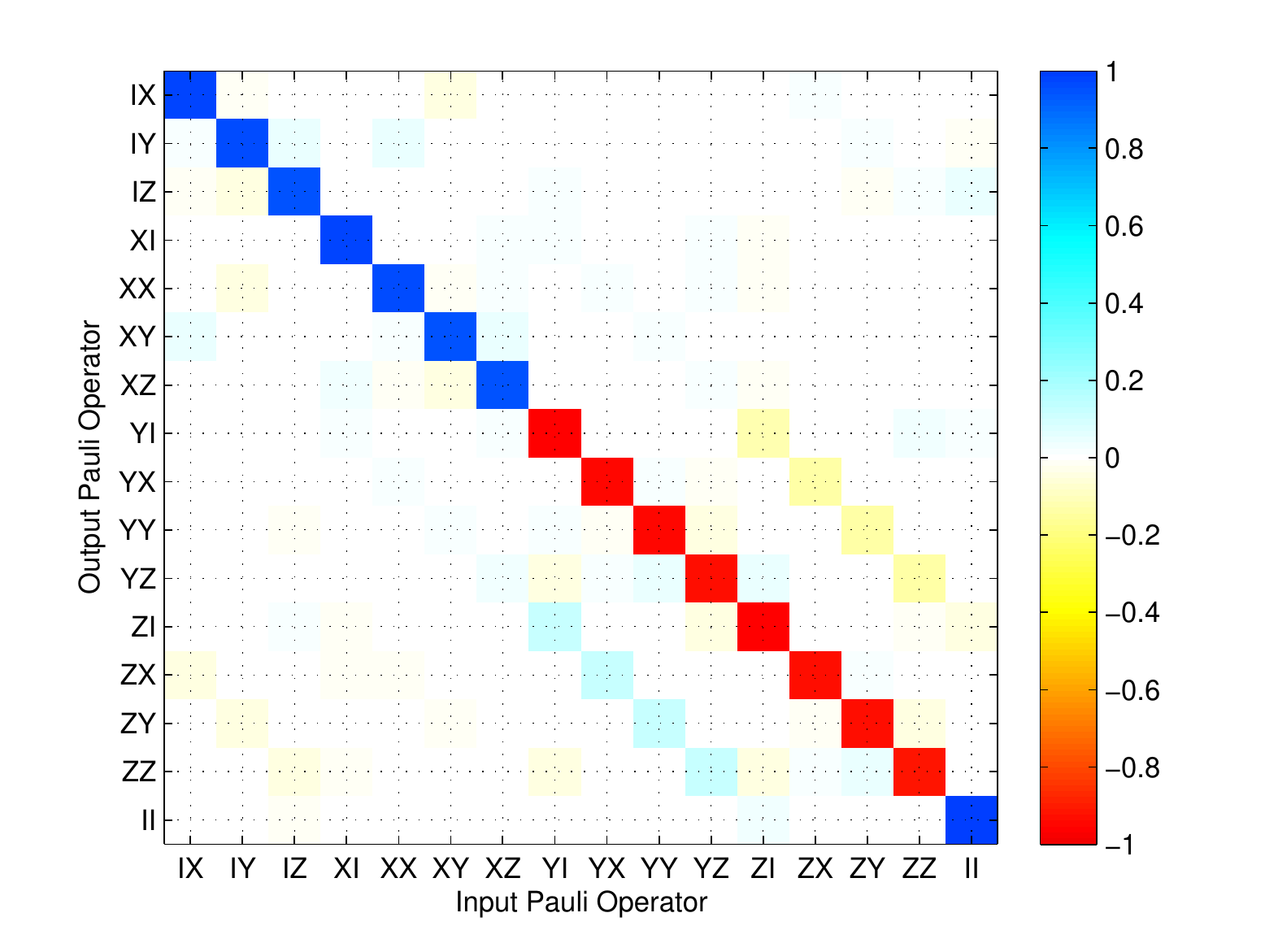}&\includegraphics[width=0.4\textwidth]{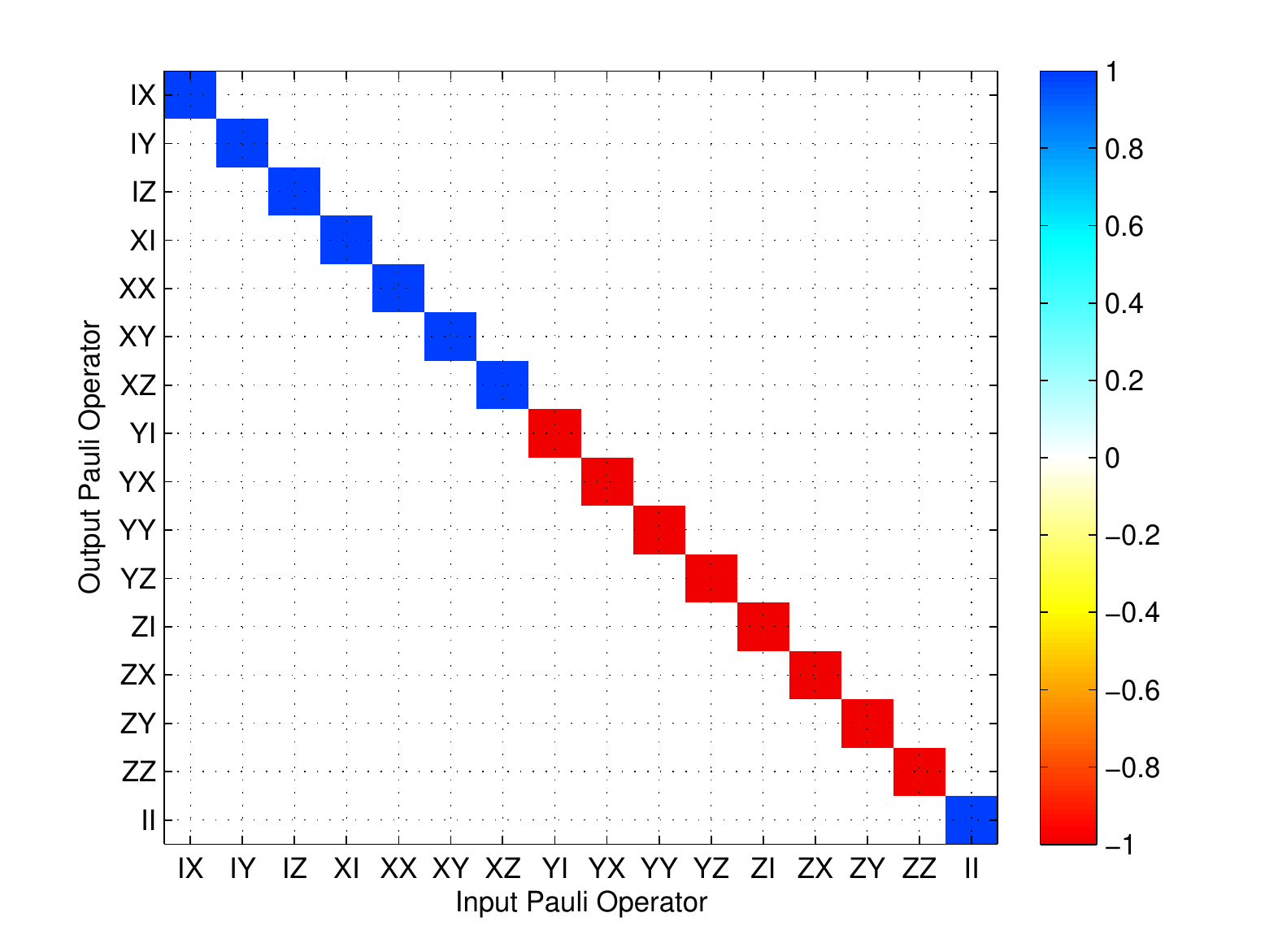} \\
Maximum likelihood & Theory\\
\hline
\end{tabular}
\caption{A comparison of the theoretical and experimentally reconstructed $\mathcal{R}$ matrices for the gates$I\otimes X_{\pi/8}$,  $I\otimes X_{\pi/4}$, $I\otimes X_{\pi/2}$, $I\otimes X_{\pi}$.}\label{fig:rauliA}
\end{figure*}

\begin{figure*}
\centering
\begin{tabular}{|c|c|}\hline\\
\includegraphics[width=0.4\textwidth]{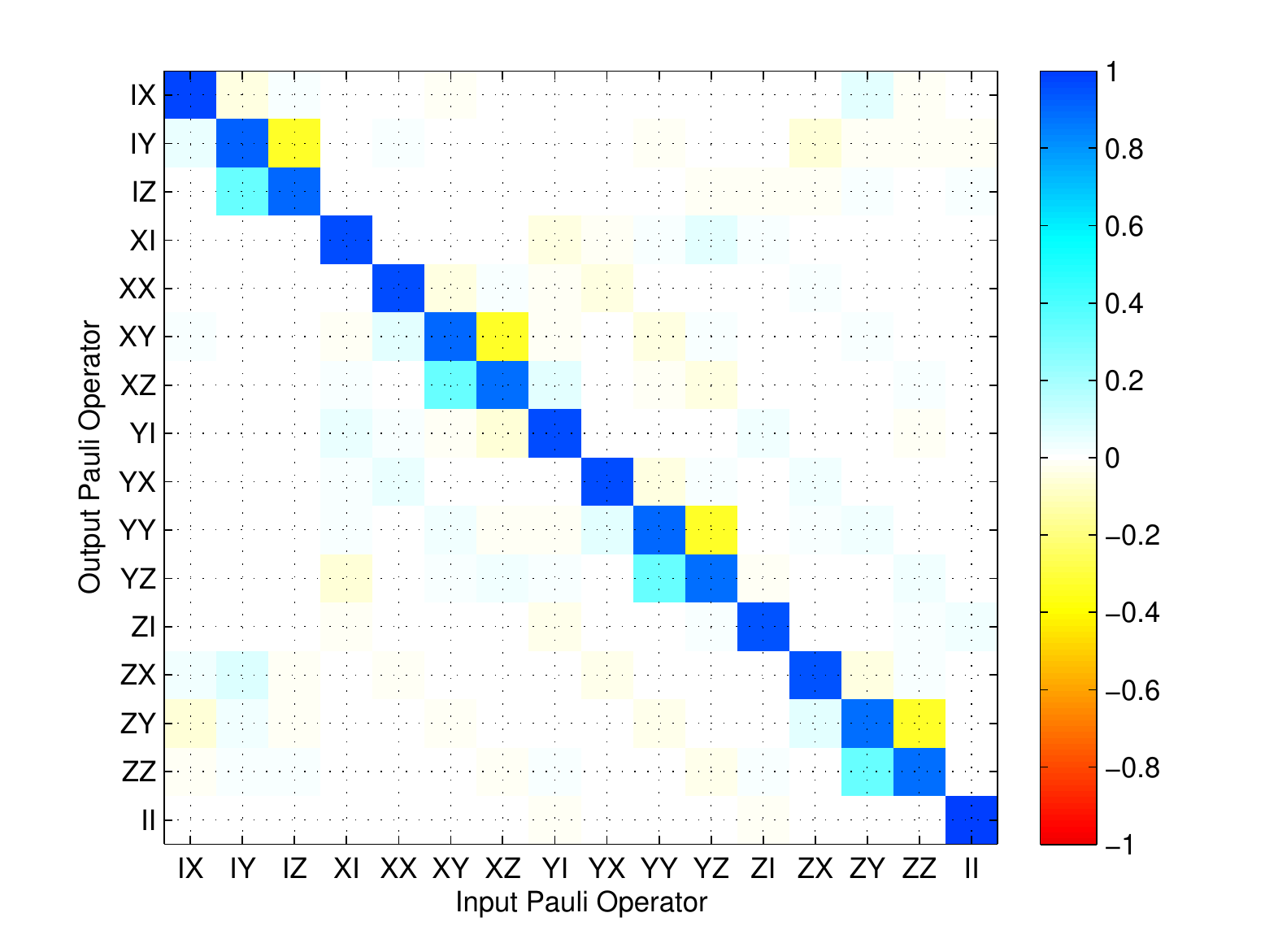}  &\includegraphics[width=0.4\textwidth]{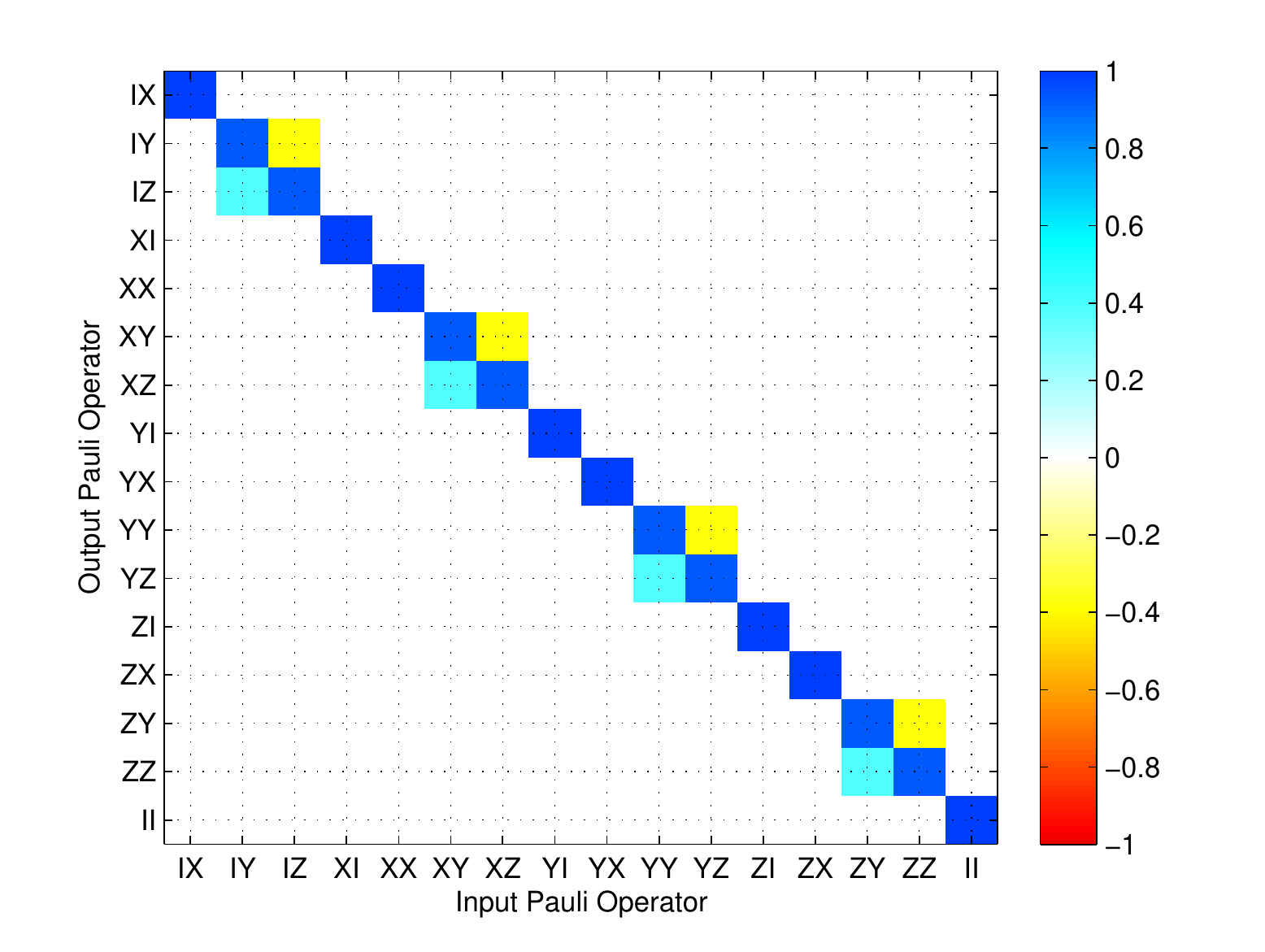}\\
\includegraphics[width=0.4\textwidth]{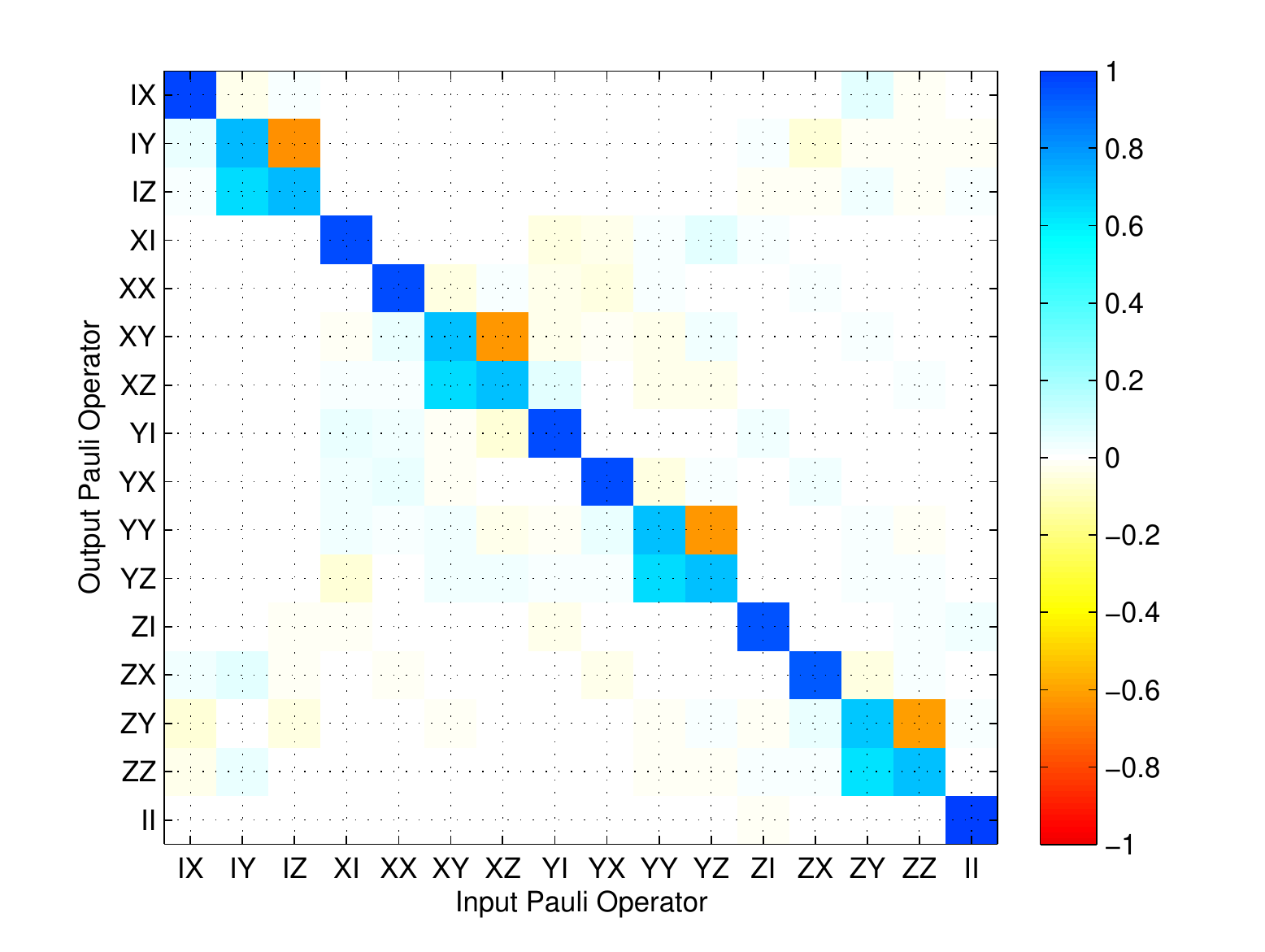}&\includegraphics[width=0.4\textwidth]{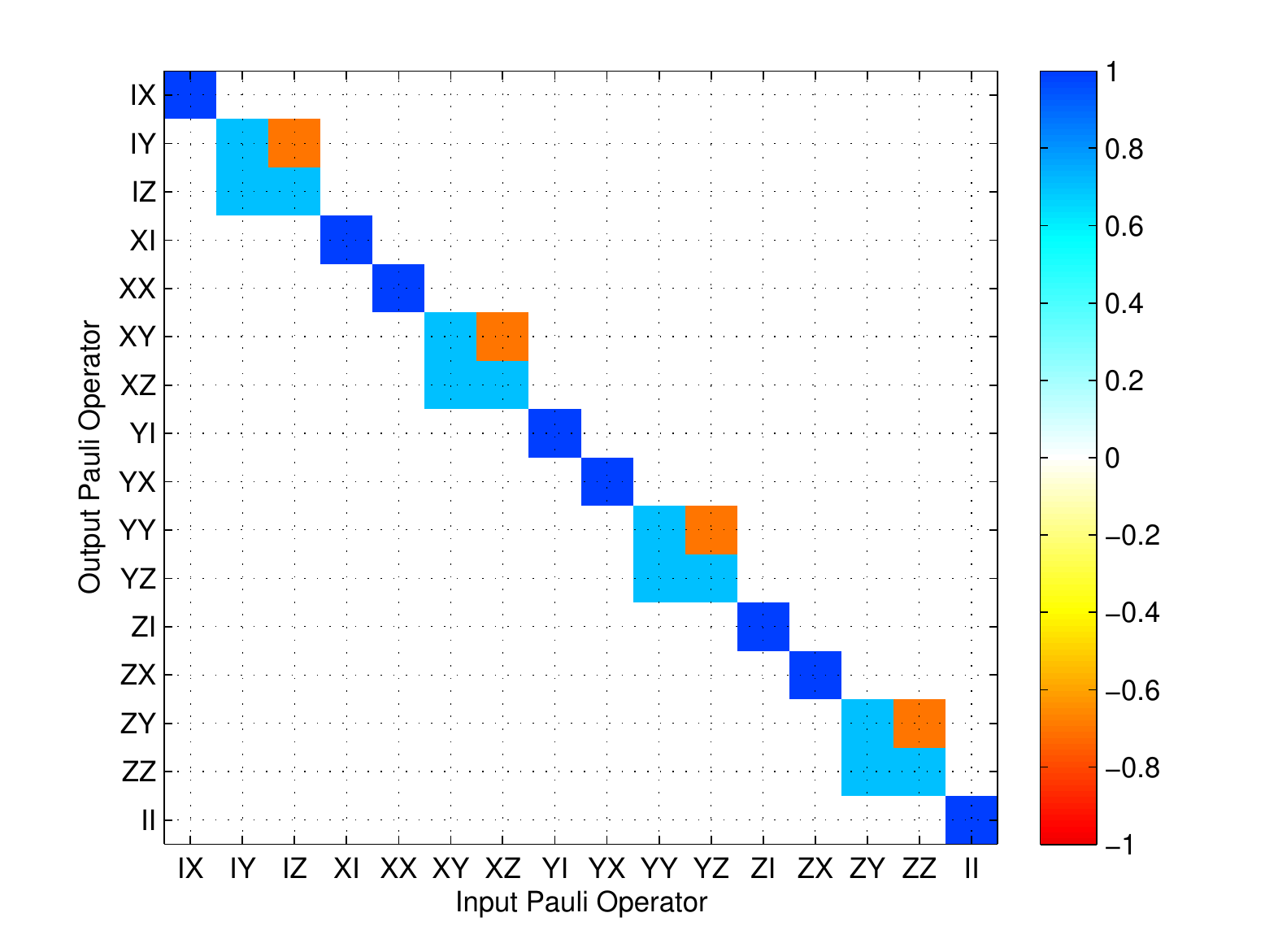}\\
\includegraphics[width=0.4\textwidth]{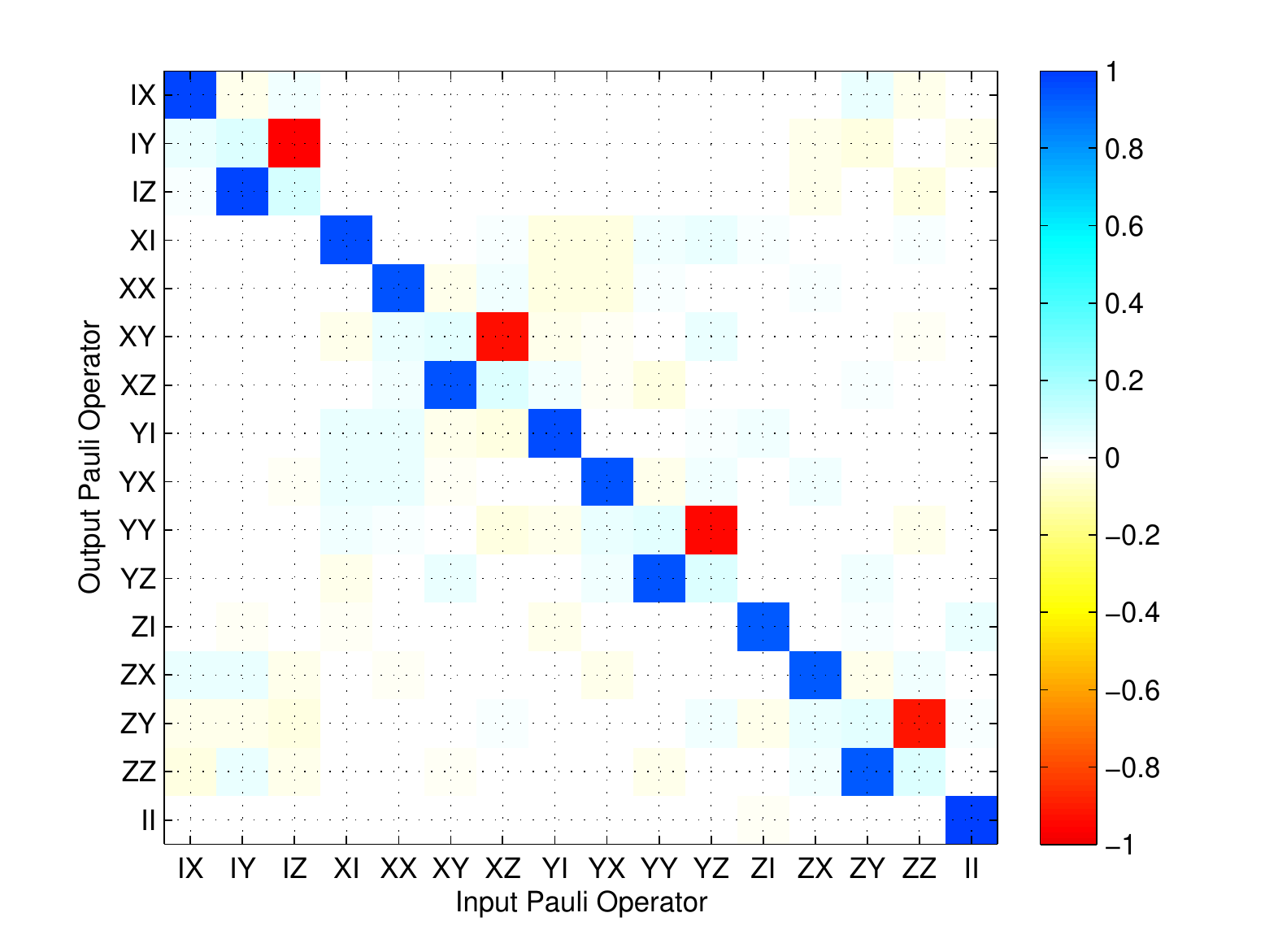} &\includegraphics[width=0.4\textwidth]{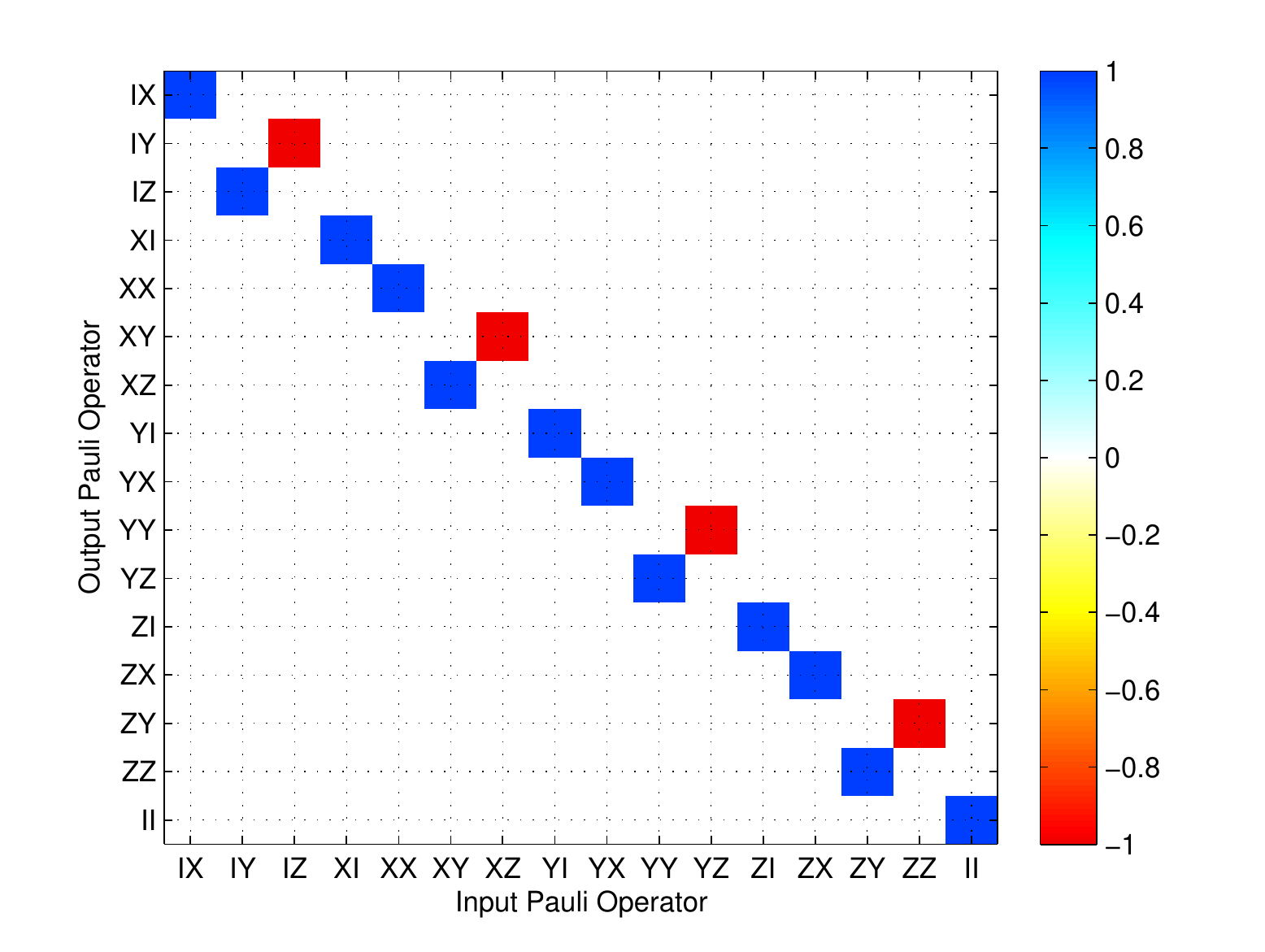}\\
\includegraphics[width=0.4\textwidth]{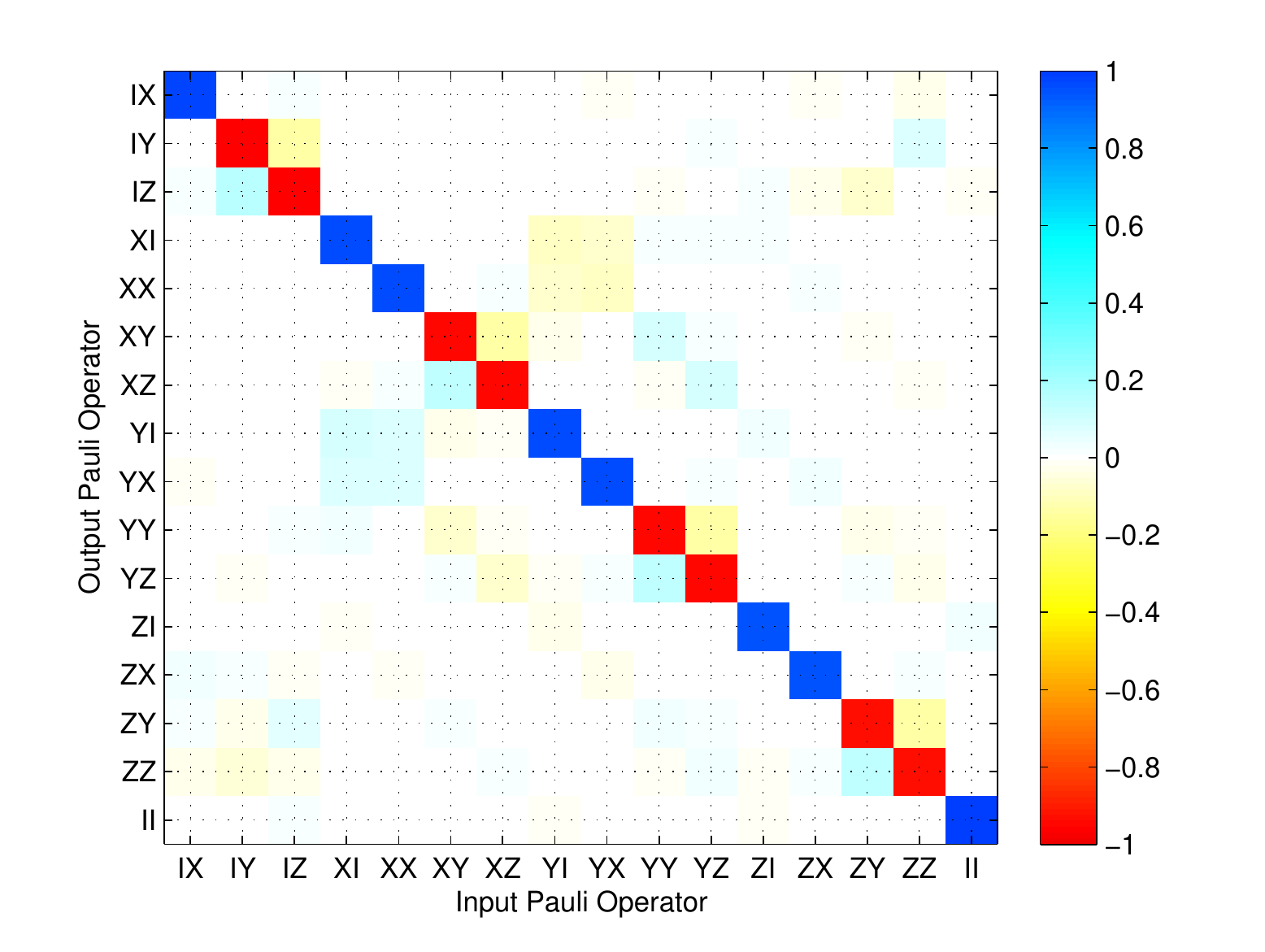} &\includegraphics[width=0.4\textwidth]{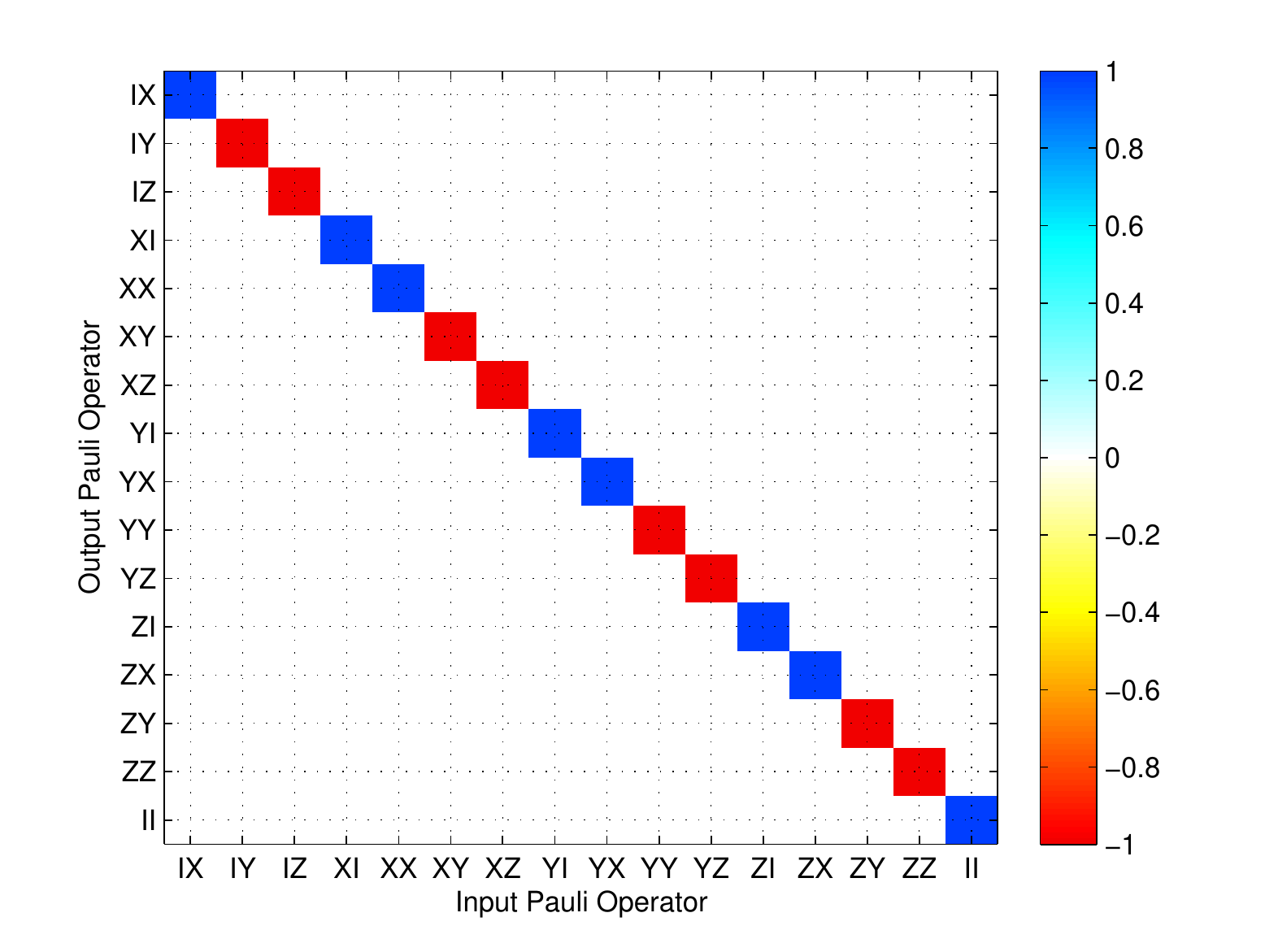} \\
Maximum likelihood & Theory\\
\hline
\end{tabular}
\caption{A comparison of the theoretical and experimentally reconstructed $\mathcal{R}$ matrices for the gates $X_{\pi/8}\otimes I$, $X_{\pi/4}\otimes I$, $X_{\pi/2}\otimes I$, $X_{\pi}\otimes I$.}\label{fig:rauliB}
\end{figure*}

\begin{figure*}
\centering
\begin{tabular}{|c|c|}\hline\\
\includegraphics[width=0.4\textwidth]{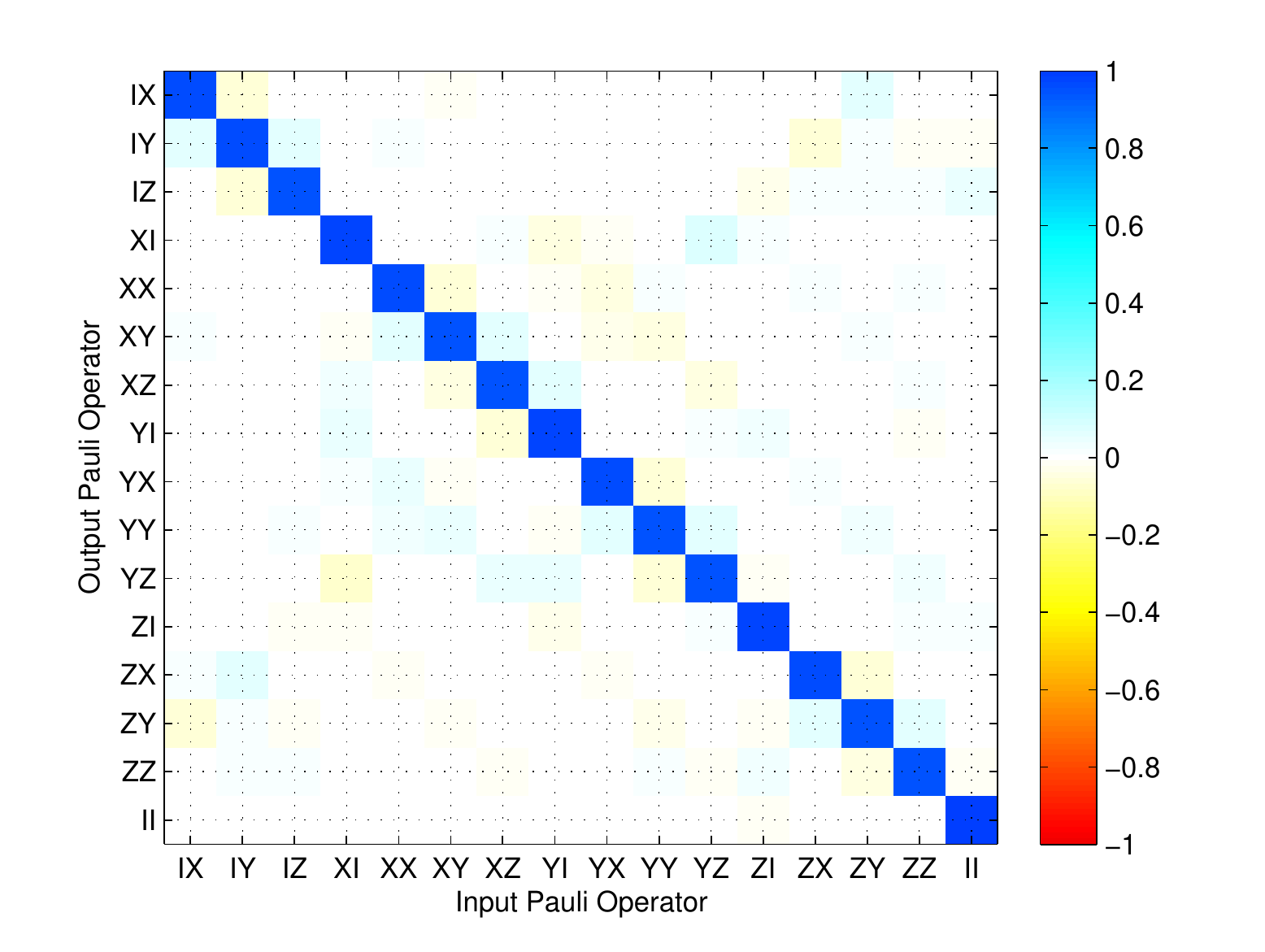} &\includegraphics[width=0.4\textwidth]{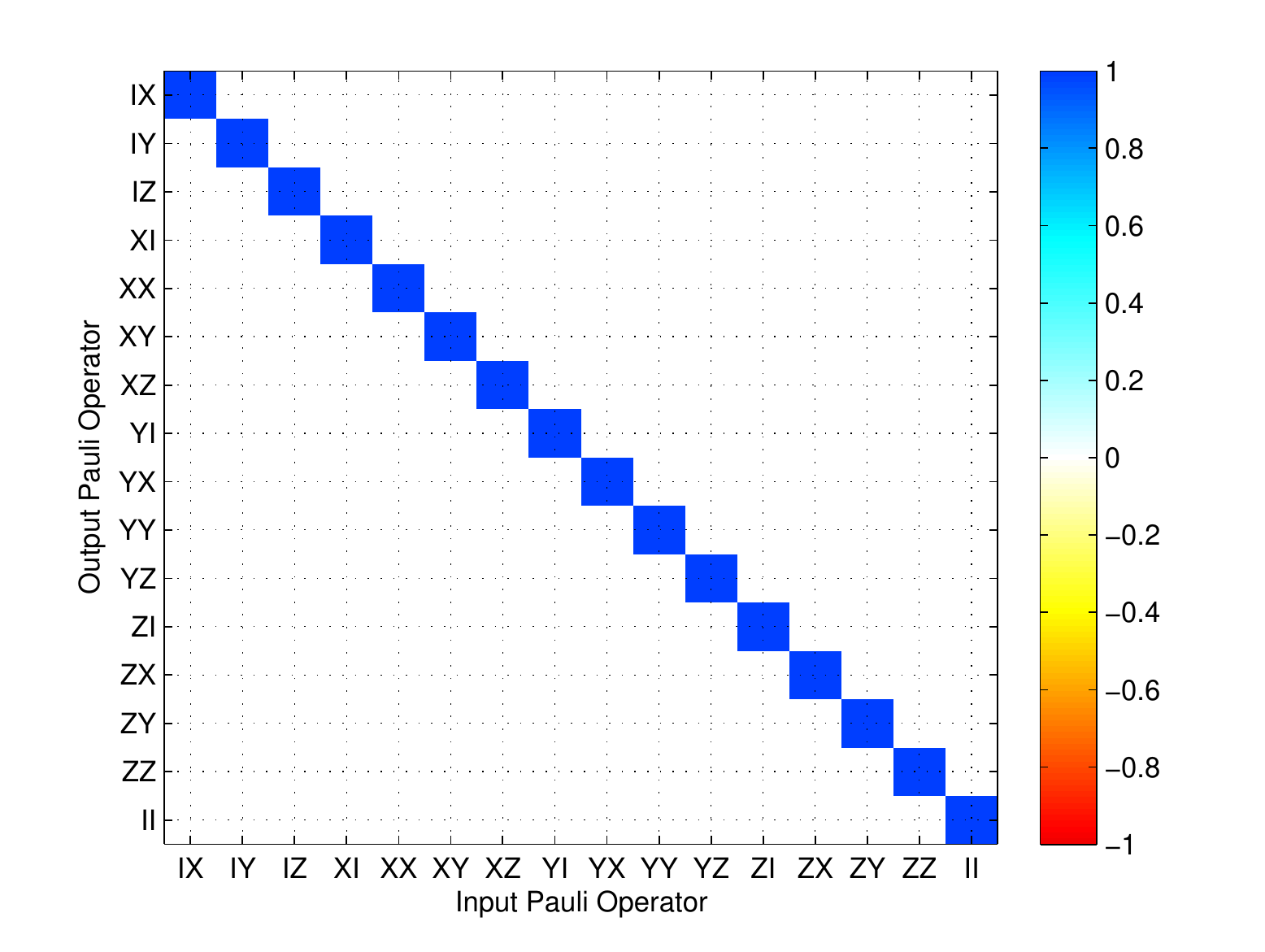}\\
\includegraphics[width=0.4\textwidth]{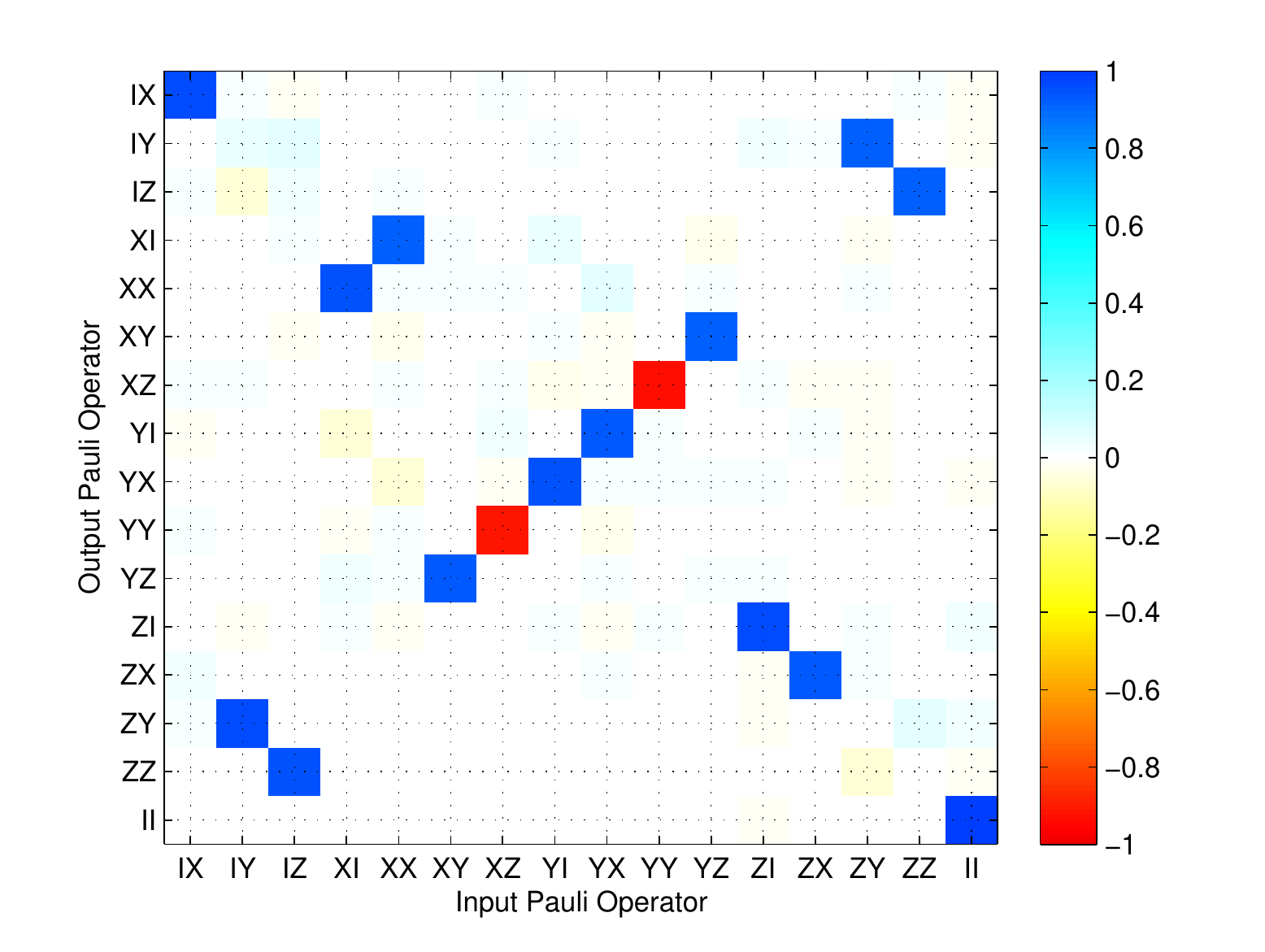} &\includegraphics[width=0.4\textwidth]{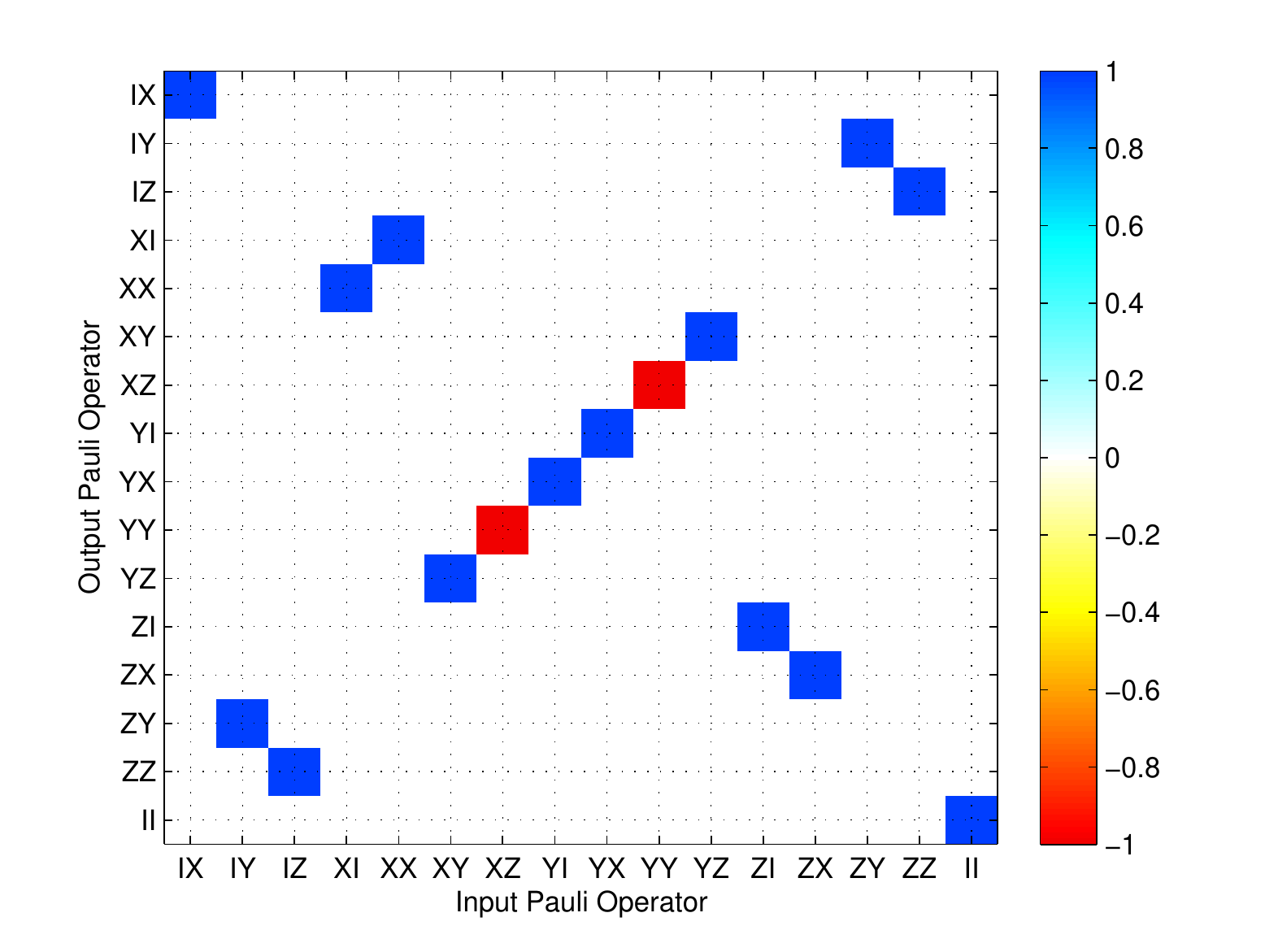}\\
\includegraphics[width=0.4\textwidth]{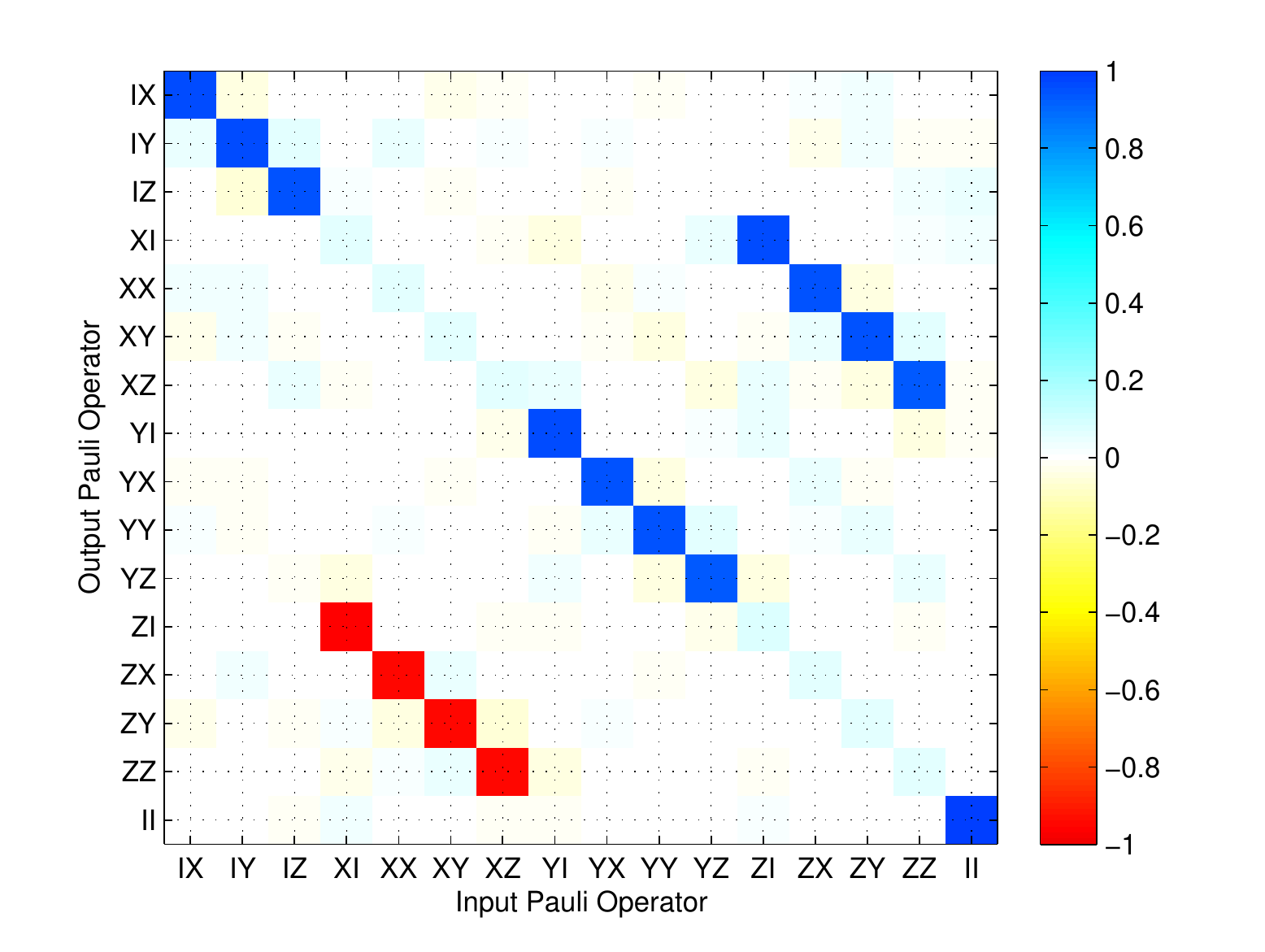} &\includegraphics[width=0.4\textwidth]{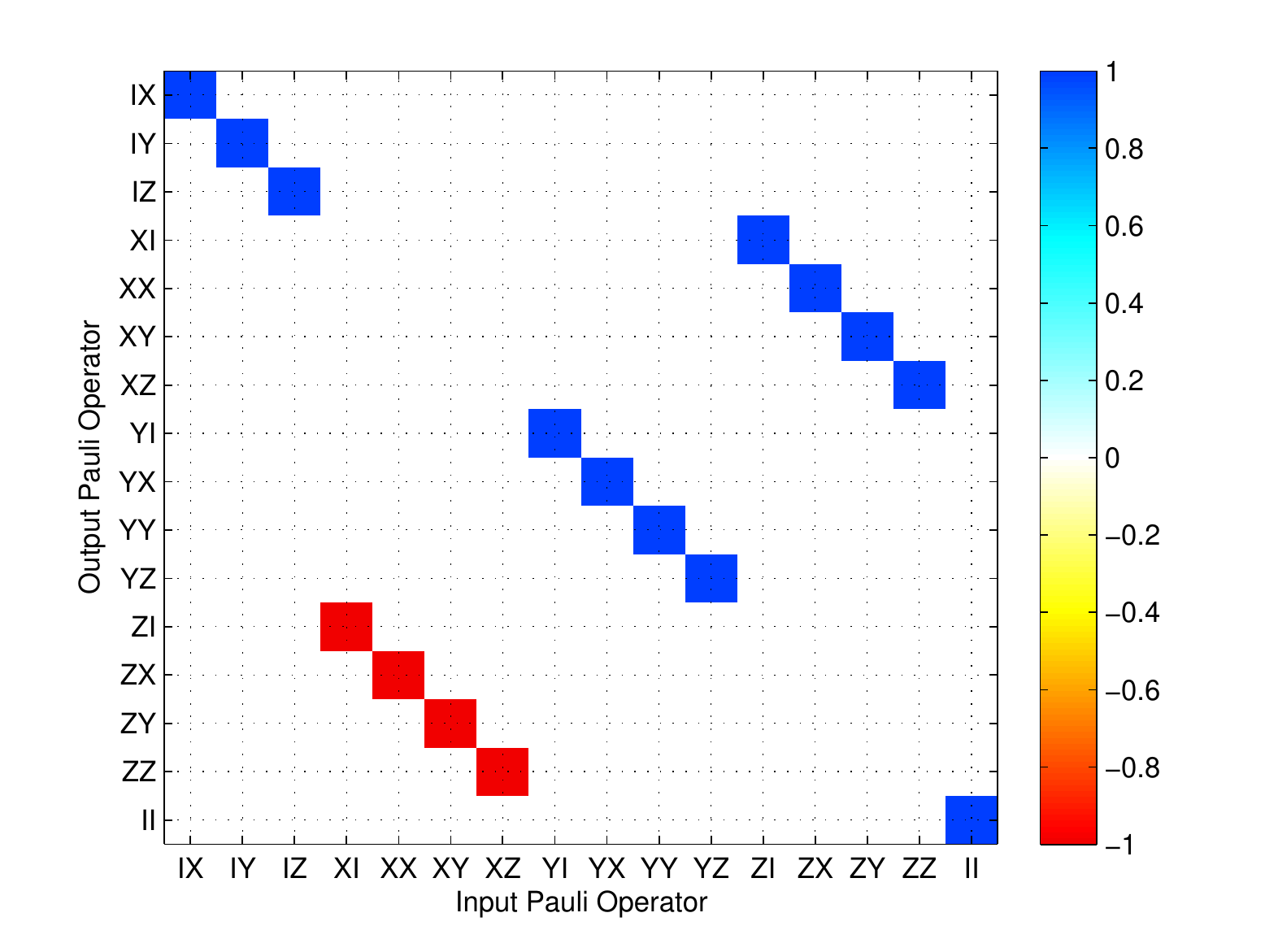}\\
\includegraphics[width=0.4\textwidth]{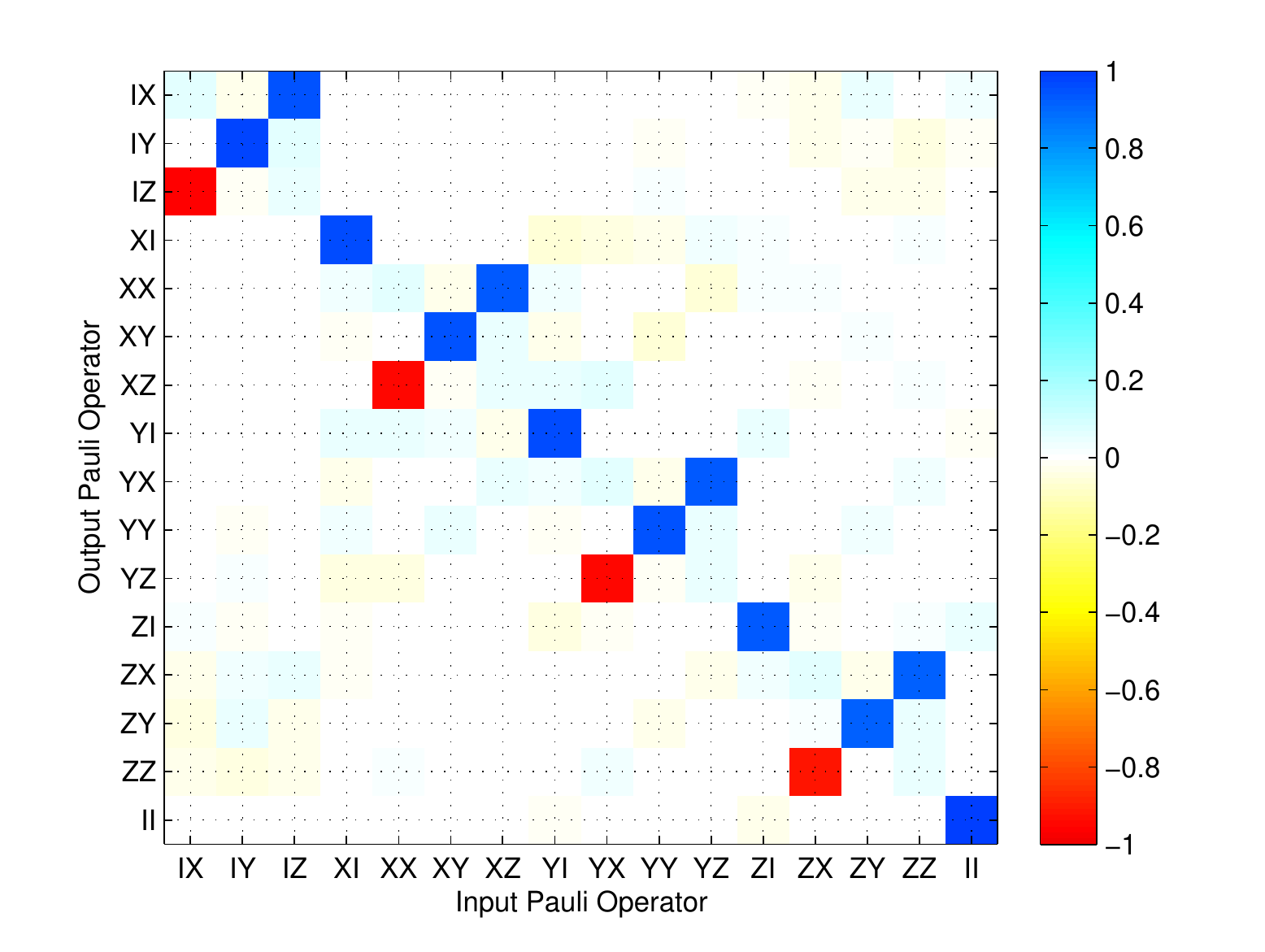}&\includegraphics[width=0.4\textwidth]{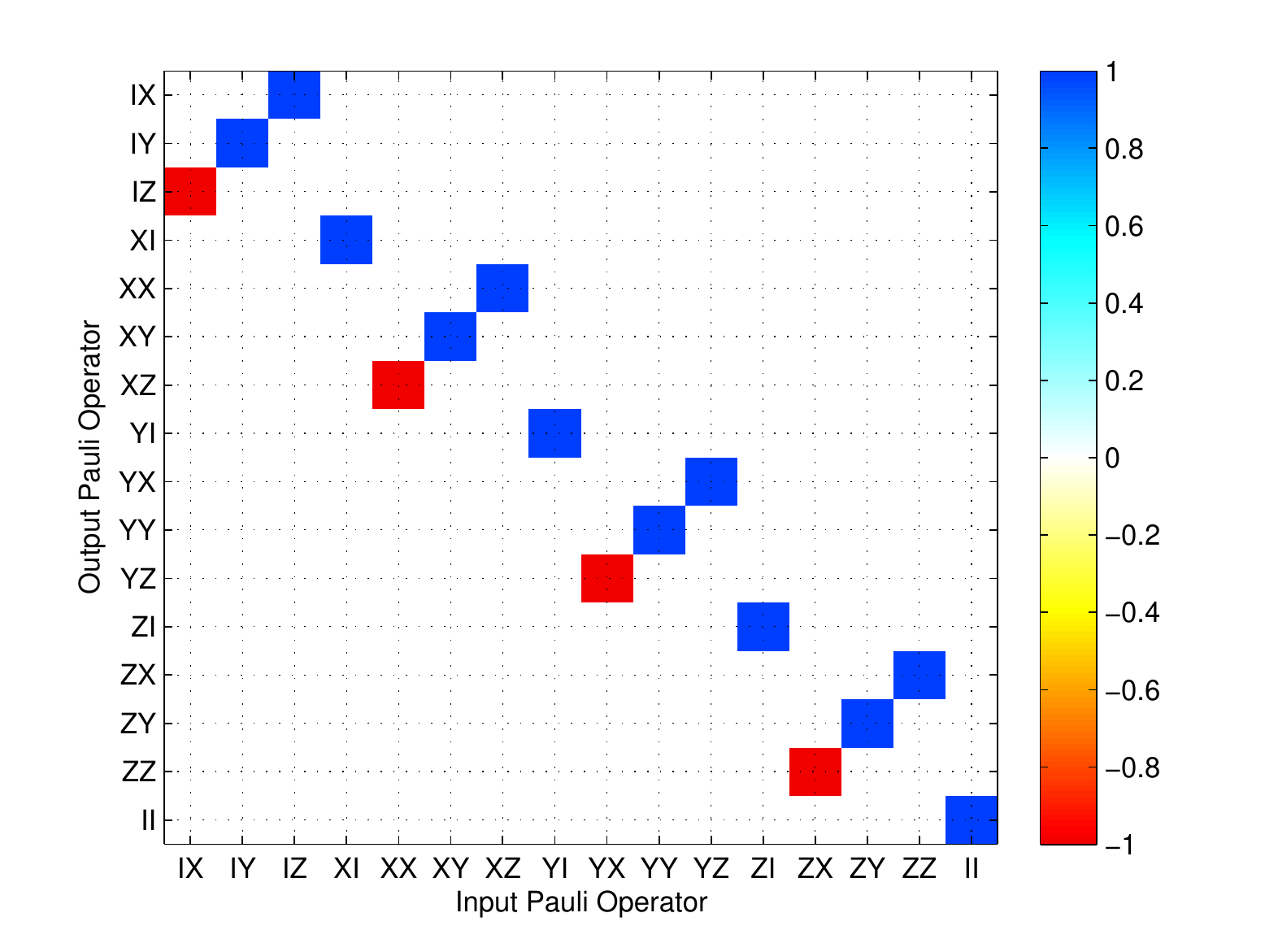}  \\
Maximum likelihood & Theory\\
\hline
\end{tabular}
\caption{A comparison of the theoretical and experimentally reconstructed $\mathcal{R}$ matrices for the gates $I\otimes I$, $\mathrm{CNOT}$, $Y_{\pi/2}\otimes I$ and $I\otimes Y_{\pi/2}$.}\label{fig:rauliC}
\end{figure*}

The average gate fidelity can be found by \cite{Nielsen2002} 
\begin{equation}F_g = \frac{d F_p +1}{d+1}\end{equation}
where the process fidelity is given by 
\begin{equation}
F_p = \mathrm{Tr}[\rho_\mathrm{ideal}\rho_\mathrm{exp}] = \mathrm{Tr}[\mathcal{R}_\mathrm{ideal}^T \mathcal{R}_\mathrm{exp}]/d^2,
 \end{equation} 
assuming that the ideal map is a unitary map.

\section{Quantum process tomography}

\subsection{The Likelihood function}

Measurements in quantum mechanics can be represented by positive operator
valued measurements (POVMs).  Mathematically, a POVM is
a decomposition of the identity into positive semidefinite Hermitian  
operators $\sum_{k=1}^n E_k = I$.  The probability of each outcome $k$
when state $\rho$ is measured is $\mathrm{Tr} (E_k \rho)$.  In the continuous limit
we have instead $\int d m E_m= I$ and the probability density for measurement
outcome $m$ is $p(m)=\mathrm{Tr} (E_m \rho)$.

For a measurement of observable $M=\sum_i m_i |i\rangle\langle i|$ with Gaussian noise of variance $v$ (as common with real amplifiers) the POVM is
\begin{equation}\label{Eis}
E_m=\frac{1}{\sqrt{2 \pi v}}\sum_i e^{-(m -m_i)^2/(2 v)}|i\rangle\langle i|
\end{equation} and the probability density  for measurement outcome $m$ is
\begin{equation}
P(m|\rho)=\frac{1}{\sqrt{2 \pi v}}\sum_i e^{-(m -m_i)^2/(2 v)}P(i|\rho).
\end{equation}
If we sample from this distribution $N$ times and are interested in  $\tilde m =\sum_n m^{(n)}/N$ then the central limit theorem asserts that as $N\rightarrow\infty$, the distribution of $\tilde m$ approximates a normal distribution with mean $\mathrm{Tr}[M\rho]$ and variance $v/N$,
  \begin{equation}
P(\tilde m|\rho)=\frac{1}{\sqrt{2 \pi v/N}} e^{-( \tilde m -\mathrm{Tr}[M\rho])^2/(2 v/N)}.
\end{equation}

To have complete information about the quantum operations we need to measure a set of linear independent observables and prepare linearly independent states. To achieve this we assume that the initial state can be reliably be prepared in the ground state and the measurement operator ($M$) is diagonal. Experimentally, from applications of $\{I, X_\pi\}^{\otimes 2}$ pulses we reconstruct the outcomes of the measurement $M$. For this work we find $M_{00} =0.0035$,  $M_{01} = 0.0196$, $M_{10} =0.0302$, and $M_{11} = 0.0323$ with a standard deviation of $\sqrt{v}= 0.0143$ (all units $\mu V$).  Having characterized the initial state and the measurement operator, we can completely determine $\mathcal{R}$ from data obtained for the applications of the pulses $\{ I, X_\pi , X_{\pm\pi/2}, Y_{\pm\pi/2}\}^{\otimes2}$ before and after the operation to be analyzed. 

Defining $\bar m_{ij}$ as the measurement result for the $i^\mathrm{th}$ preparation and $j^\mathrm{th}$ measurement we write the Likelihood function 
\begin{equation}\mathcal{L}(\Lambda) = \prod_{i,j} P( \bar m_{ij}|\Lambda) =  \prod_{i,j} \frac{e^{-( m_{ij}- \bar m_{ij} )^2/(2v_j/N)}}{\sqrt{2\pi v_j/N}} \end{equation}
where 
\begin{equation} m_{ij} = \mathrm{Tr}[M_j \Lambda(\rho_i)]= \frac{1}{d}\sum_{mn} \mathrm{Tr}[\rho_i P_n]\mathrm{Tr}[M_j P_m]\mathcal{R}_{mn}.  \label{results}\end{equation} 
Here we have allowed the measurements to have different variances. Maximizing this likelihood function is equivalent to minimizing the negative log-likelihood function,
\begin{equation}
{\cal L}_{\log}=\sum_{ij} [ m_{ij}- \bar m_{ij}]^2/v_i,
\end{equation}
and so then the problem of Quantum Process Tomography (QPT) becomes 
\begin{equation}
\begin{split}
 \mathrm{minimize} &~ \vec{\delta}(\Lambda)^TV^{-1} \vec{\delta}(\Lambda)\\
 \mathrm{subject~to} &~ \rho_\Lambda \geq 0 \label{QPT1}.
\end{split}
\end{equation}
Here $\delta(\Lambda)=  m_{ij}- \bar m_{ij} $ and $V$ is the covariance matrix.

\begin{table*}
    \begin{ruledtabular}
    \begin{tabular}{|l|c|c|c|c|c|c|c|c|c|c|}
   	Gate & $F_p$ & $F_g$ & $\Delta F_g$ & $R_{I,I}$ & $F_\mathrm{pure}$ & $\lambda_\mathrm{max}$ & $\sum_{\lambda<0}\lambda$ & $0.5\|\rho_\mathrm{MLE}-\rho_\mathrm{ideal}\|_2$  & $0.5\|\rho_\mathrm{MLE}-\rho_\mathrm{data}\|_2$ & $0.5\|\rho_\mathrm{data}-\rho_\mathrm{ideal}\|_2$  \\
 &  & & ($\times10^{-4}$) & & & & &  &  &   \\
    \hline
   	$I\otimes I$ &0.9614& 0.9691 & 3.6 &1.0043 & 0.9954 & 0.9656& -0.082 & 0.050 & 0.036& 0.062  \\ 
	$X_{\pi}\otimes I$ &0.9522& 0.9618 & 4.3 &1.0013&  0.9935 &0.9584&-0.079 & 0.061 & 0.029 & 0.063 \\ 
	$X_{\pi/2}\otimes I$ &0.9525& 0.9620 & 5.2 &1.0009& 0.9955 &0.9568&-0.083 & 0.053 & 0.030 &  0.059 \\ 
	$Y_{\pi/2}\otimes I$ &0.9526& 0.9621 & 5.3 &1.0018&  0.9956 &0.9568&-0.119 & 0.052 & 0.046 & 0.070 \\ 
	$X_{\pi/4}\otimes I$ &0.9608& 0.9687 & 5.5 &0.9975&  0.9962 &0.9644&-0.096 & 0.047 & 0.038 & 0.059 \\ 
	$X_{\pi/8}\otimes I$ &0.9561& 0.9649 & 5.2 &0.9980&  0.9962  &0.9597&-0.088 & 0.048 & 0.038 & 0.062 \\ 
	$I\otimes X_{\pi}$  &0.9537& 0.9629 & 4.4 &0.9987&  0.9906 &0.9627&-0.090 & 0.071 & 0.033 & 0.076 \\ 
	$I\otimes X_{\pi/2}$ &0.9497& 0.9597 & 3.9 &0.9954&  0.9955 &0.9539&-0.081 & 0.054 & 0.031 & 0.062 \\ 
	$I\otimes Y_{\pi/2}$ &0.9461& 0.9569 & 4.6 &1.0006&  0.9961 &0.9498&-0.104 & 0.052 & 0.040 & 0.065 \\ 
	$I\otimes X_{\pi/4}$ &0.9554& 0.9644 & 4.6 &0.9970& 0.9963 &0.9590 &-0.091 & 0.048 & 0.035 & 0.058 \\ 
	$I\otimes X_{\pi/8}$ &0.9583& 0.9666 & 6.0 &0.9964& 0.9968 &0.9613&-0.106  & 0.045 & 0.042 & 0.061 \\ 
	CNOT &0.9384 & 0.9507 & 6.5 &1.0031 & 0.9986 &0.9396&-0.105 & 0.044 & 0.035 & 0.055
 \end{tabular}
    \end{ruledtabular}
    \caption{A compilation of various error measure and measures of physicality for our universal set of quantum gates.  The columns, from left to right, contain: the name of the gate, the process fidelity, the gate fidelity, the error in gate fidelity found from bootstrapping, the value of the Pauli transfer matrix from the identity to the identity (should be unity for a physical map), the purified fidelity, the maximum eigenvalue of the reconstructed state, the sum of the negative eigenvalues of the reconstructed state obtained by simple matrix inversion and finally three two norm distances between the pairwise combinations of the ideal gate ($\rho_\mathrm{ideal}$), the physical maximum likelihood estimate ($\rho_\mathrm{MLE}$) and the unphysical gate obtained by inverting the data ($\rho_\mathrm{data}$). }\label{fidtab}
\end{table*}

\subsection{Maximum Likelihood estimation using a semidefinite program}

A general semidefinite program (SDP) can be expressed in the standard form:
\begin{equation}
\begin{split}
 \mathrm{minimize} &~\vec{b}^\mathrm{T}\vec{x}\\
 \mathrm{subject~to} &~ F_0 + \sum_i F_i x_i \geq 0
\end{split}
\end{equation}
where the $n$-element real vector $\vec{x}$ contains the variables to be optimized with respect to the linear objective function described by an $n$-element real vector $\vec{b}$ and constrained by the constraints are
described by a set of $m$-by-$m$ Hermitian matrices $F_0$, $F_i$ where $i = 1, ...n$.

Since the objective function of Eq.~\eqref{QPT1} is quadratic, it is not in the standard form. In order to rectify this we introduce a slack variable $t$ defined by  $t-\vec{\delta}(\Lambda)^TV^{-1} \vec{\delta}(\Lambda)\geq0$.   By the Schur complement the previous inequality is satisfied if and only if 
\begin{equation}
Z = \left(\begin{matrix} t&\vec{\delta}\\ \vec{\delta}^T&V \end{matrix}\right)\geq 0
\end{equation} and $V\geq0$. Therefore, we can express the QPT problem as 
\begin{equation}
\begin{split}
 \mathrm{minimize} &~\vec{b}^\mathrm{T}\vec{y}\\
 \mathrm{subject~to} &~ A=\left(\begin{matrix} Z&0\\ 0&\rho_\Lambda \end{matrix}\right)\geq 0
\end{split}
\end{equation} where $b=\{1,0_{d^4}\}$ and $\vec{y}=\{t,\vec{r}\}$ with $\vec{r}$ being the column-major vector representation of $\mathcal{R}$.  Although there are many numerical libraries for solving semidefinite programs in this standard form, we choose to use SeDuMi \cite{sedumi}.

In Fig.~\ref{fig:rauliA} - \ref{fig:rauliC} we show the experimental $\mathcal{R}$ matrix after applying MLE as well as the theoretical $\mathcal{R}$.  These are plotted for the measured operations $\{ I, X_{\pi/8}, X_{\pi/4}, X_{\pi/2},  X_{\pi}, Y_{\pi/2}\}^{\otimes2}$ and CNOT. The corresponding gate fidelities are shown in Table~\ref{fidtab}, and are all above $0.95$.

\section{Error estimation} 

\subsection{Statistical analysis of errors}

The conceptually simplest way to obtain a confidence region on our QPT extracted error rates would be to calculate
\begin{equation}
p = \int_{\mathcal{S}} {\rm d} \rho_\Lambda P(\rho_\Lambda | {\rm data}) =  \frac{\int_{\mathcal{S}} {\rm d} \rho_\Lambda \mathcal{L}(\rho_\Lambda)}{\int_{\mathcal{V}} {\rm d} \rho_\Lambda \mathcal{L}(\rho_\Lambda)} \label{eq:confidence}
\end{equation}   
where we interpret $\mathcal{S}$ as the region that contains the actual state with probability $p$ ($\mathcal{V}$ is the entire space of Choi matrices).  Even if we model the noise on our measurements as Gaussian, constraining the domain of the likelihood function to the set of positive semidefinite matrices makes this integration extremely difficult.

The simplest way to bound the error due to any noise process would be to take many independent realizations of an experiment.  However, given reduced measurement fidelities, it can be costly to run the experiment a large number of times, and instead we can use a bootstrapping method to approximate a new realization of the noise on an existing data set.  Given the measurement record $\bar m_{ij} $ we can simulate 
\begin{equation}
 \tilde m^{(k)}_{ij} =\bar m_{ij} + \sqrt{v_j/N} W^{(k)}_{ij}.
\end{equation}  where $W^{(k)}_{ij}$ is a gaussian random variable of mean 0 and variance 1 and $N=10000$. From this new measurement record we can compute an ensemble of maximum likelihood estimates $\rho^{(k)}_{\rm MLE}$.  This ensemble is not the same as averaging over different realizations (in particular the mean is offset) but the variance of the distribution serves as an upper bound on the statistical fluctuations in the estimate. In Table \ref{fidtab} the error for one standard deviation gives about $5\times10^{-4}$, which is much smaller than the measured error ($\approx 5\%$).

\subsection{Systematic errors} 
\begin{figure*}
\centering
\begin{tabular}{cc}
\includegraphics[width=0.4\textwidth]{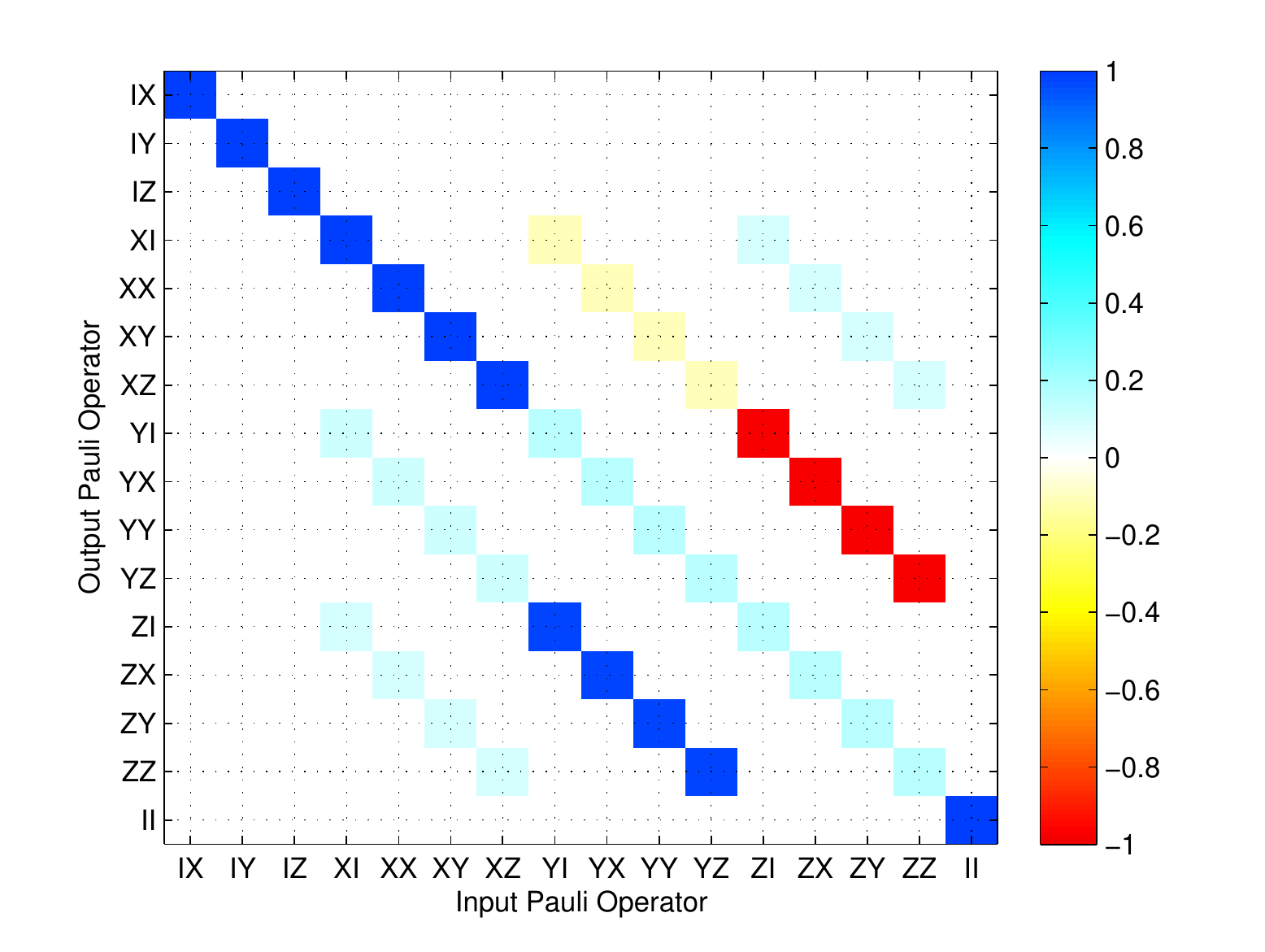} &
\includegraphics[width=0.4\textwidth]{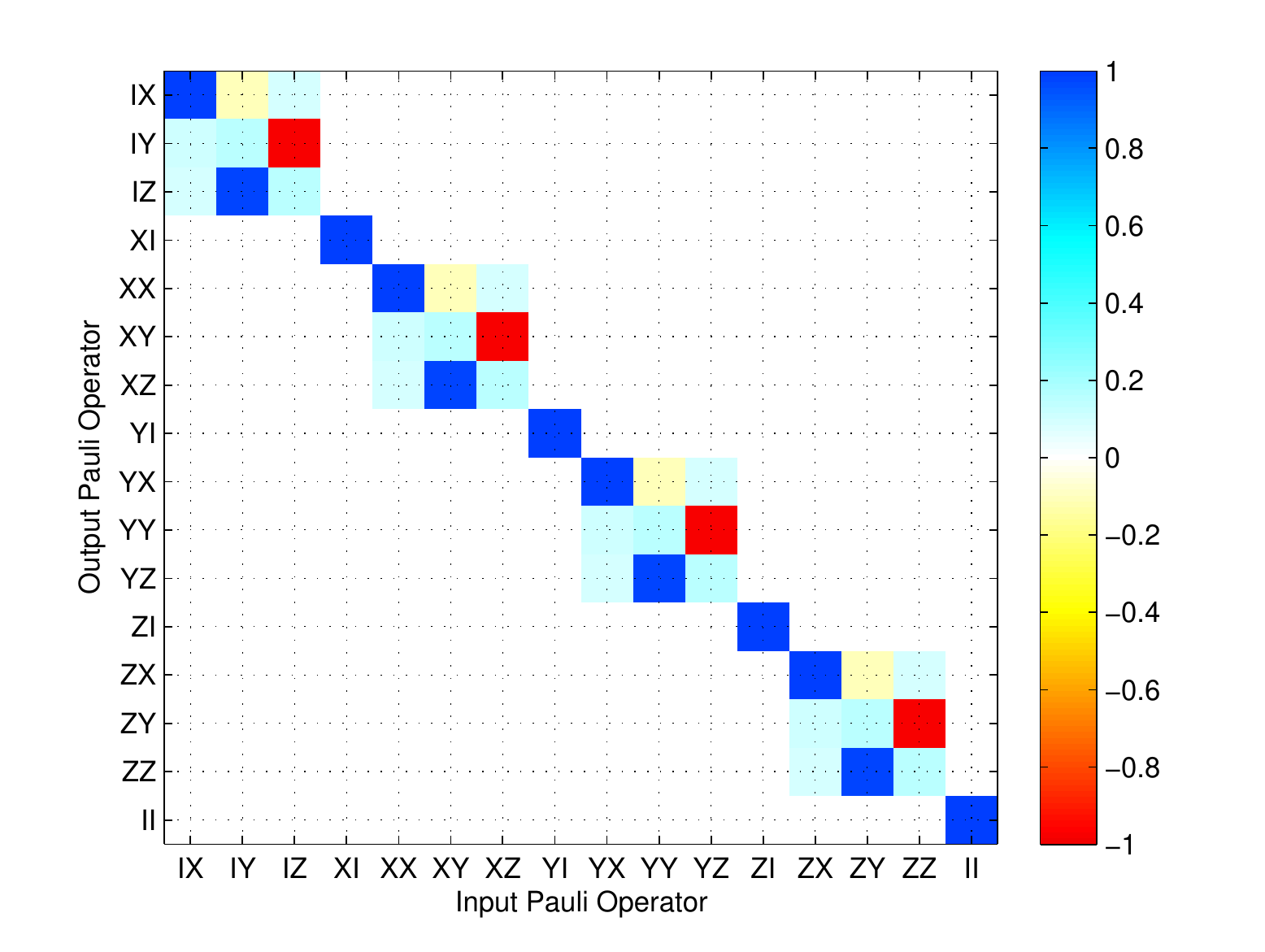} \\
(a)&(b)\\
\includegraphics[width=0.4\textwidth]{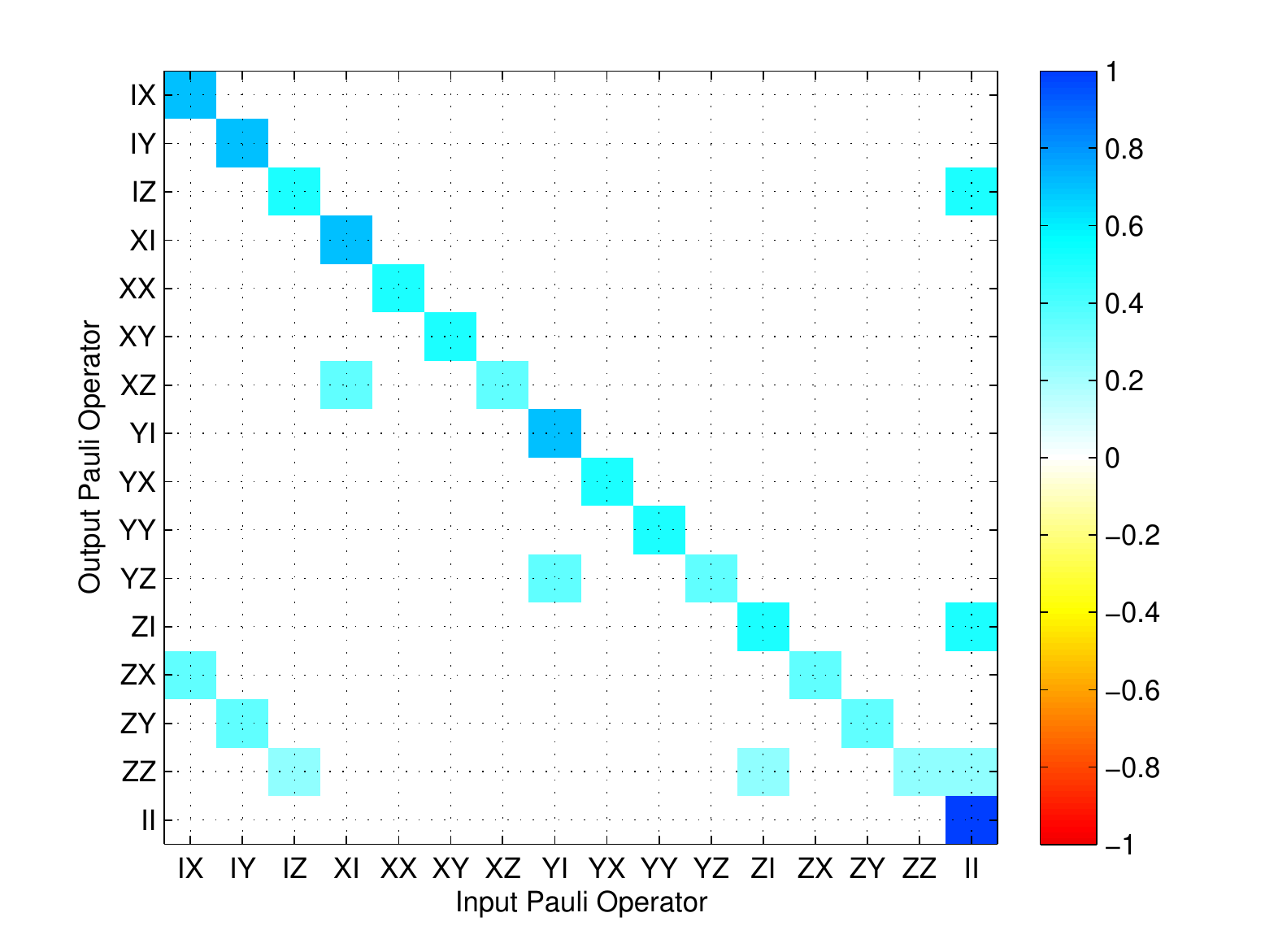}&
\includegraphics[width=0.4\textwidth]{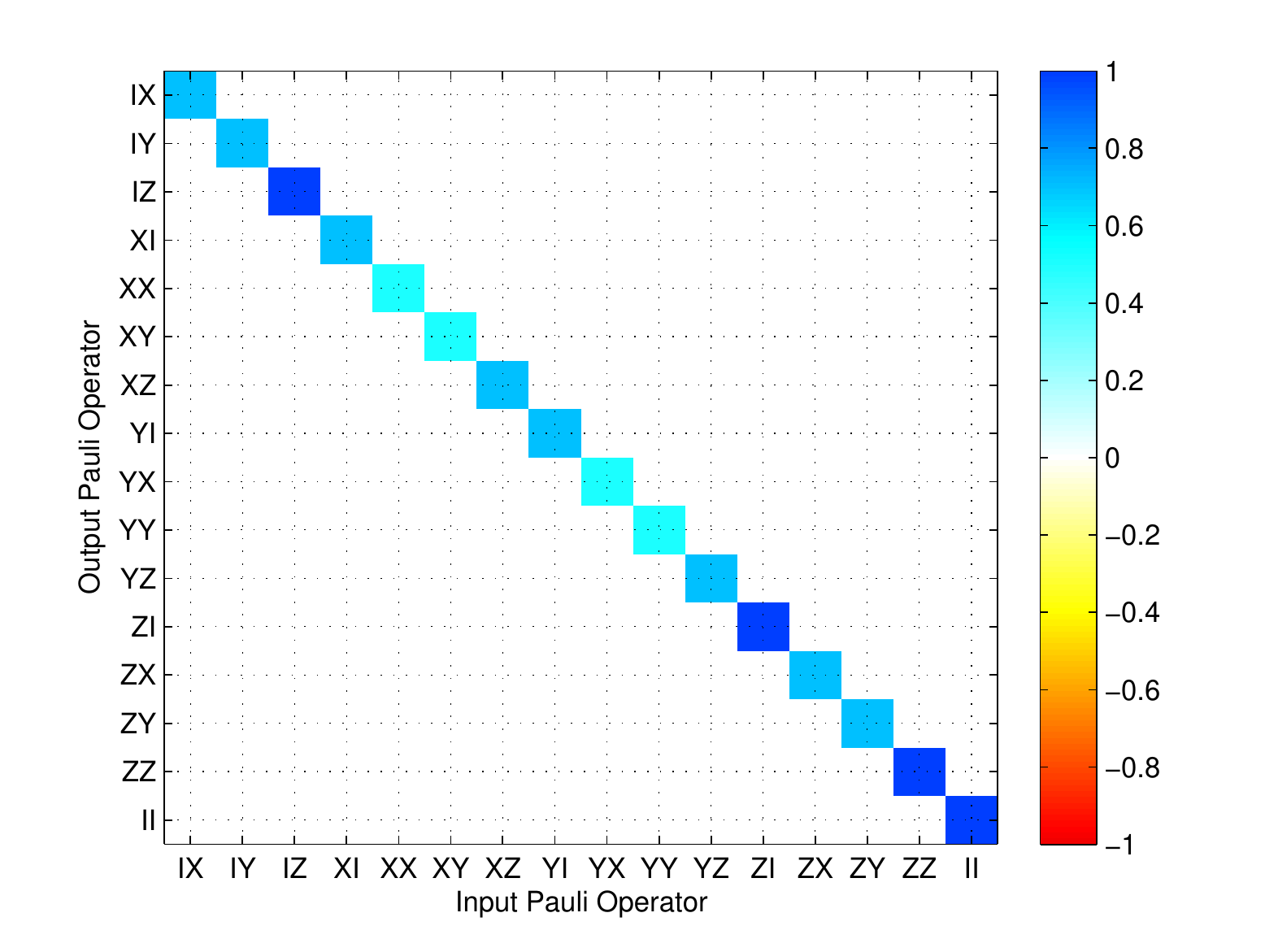}\\
(c)&(d)
\end{tabular}
\caption{The $\mathcal{R}$ matrix for (a) a $X_{\pi/2}\otimes I$ with a $1\%$ under rotation and $1\%$ phase error, (b) $I\otimes X_{\pi/2}$ with a $1\%$ under rotation and $1\%$ phase error, (c) a amplitude damping channel (with error probability of 0.5), (d) a dephasing channel (with error probability of 0.5). }\label{error}
\end{figure*}

There are a number of errors that can lead to systematic errors such as decoherence, over-under rotations, phase errors, etc. These errors can occur in both the gate being characterized as well as the gates used in the analysis and input state preparations.  Some examples of $\mathcal{R}$ maps for these types of errors are shown in Fig.~\ref{error}.  

In our experiment the coherence times were $T_1^{(1)} =8.2~\mu$s, $T_1^{(2)} =9.7~\mu$s, $T_2^{(1)} =7.1~\mu$s and $T_2^{(2)} =10.3~\mu$s and we can estimate that this would lead to an error of $1.62\%$ for the single qubit gates (total time for preparation, process and analysis pulses of 120 ns) and $2.55\%$ for the CNOT (total time 190 ns). This is much smaller than the observed error, making it likely that there are other, larger sources of error as well. 
This is confirmed by looking more closely at the $\mathcal{R}$ matrix, Fig.~\ref{fig:rauliA} and Fig.~\ref{fig:rauliB}. The features associated with amplitude damping are not present, whereas we clearly see signatures of under rotations and phase errors, see the $\mathcal{R}$ plots of the  $X_{\pi/2}\otimes I$ and $I\otimes X_{\pi/2}$ in Fig \ref{error}. We estimate that this coherent error accounts for about $1 \%$ of the total gate error in our experiment.  This is confirmed to some extent by the high values in the purified fidelity. The purified fidelity is the overlap of the largest weighted eigenvector with the ideal state ($F_\mathrm{pure} = \mathrm{Tr}[\rho_\mathrm{ideal}\rho_{\lambda_\mathrm{max}}]$. This confirms that we are making the correct unitary with $\approx 0.5-1\%$ error but we are adding on to it some non-purity conserving operations from systematic errors (the weight of the maximum eigenvalue is about 0.94-0.96). 

To have a measure of the systematic errors we reconstruct a $\mathcal{R}$ matrix without enforcing the Choi matrix to be positive semidefinite. This is done by using 
\begin{equation}
\vec{r} = (W^TW)^{-1} W^T \vec{m} 
\end{equation} where $\vec{m}$ is the  column-major vector representation of $\tilde{m}$ (measured data) and $W$ is the transfer matrix defined by Eq. \eqref{results}. To quantify how much the $\mathcal{R}$ matrix has changed due to the 
SDP, we calculate the total weight of negative eigenvalues (see Table \ref{fidtab}). Here we see that about $10\%$ of the eigenvalue weight is negative. Since rotational and phase errors in the operation pulse can not result in non-physical operations and our statistical fluctuations are small this must arise from errors in the preparation and mesaurement pulses. To estimate the error induced by this we calculate the 2-norm distance, 
\begin{equation}\|\rho_1-\rho_2\|_2 = \sqrt{\mathrm{Tr}[|\rho_1-\rho_2|]}\end{equation}
between the Choi matrices of the maximum likelihood estimate, a direct (nonphysical) inversion of the data, and the ideal map.  These results suggest that we have about a $5\%$ error from the ideal operation and the physical operations have about a $3-4\%$ error from the reconstructed. This is all consistent with the analyzing pulses having over/under rotations and phase errors which can be very loosely thought of as a mechanism that reduces the purity of the operation (add more non-zero eigenvalues to the Choi matrix).    

\begin{table}
    \begin{ruledtabular}
    \begin{tabular}{ccc}
    $\langle F_g \rangle$  & $F_g$ &  $0.5\|\rho_\mathrm{MLE}-\rho_\mathrm{ideal}\|_2$    \\
    (Tomography Pulses)  & & \\
    \hline
   	0.9434 & 0.8627&  0.2359 \\
   	0.9885  & 0.9569 & 0.1088 \\
   	0.9942 & 0.9729 & 0.0802 \\
   	0.9988 & 0.9904 & 0.0355
 \end{tabular}
    \end{ruledtabular}
    \caption{Simulations of tomography results when faulty preparation and measurement pulses are used to reconstruct a perfect identity operation.  Each simulation is averaged over ten random gate set errors. }\label{simtab}
\end{table}

One simple test of this conjecture is to simulate the tomography data for a perfect process using preparation and measurement pulses drawn from a faulty gate set.  We take the $36$ element gate set $\{I, X_\pi , X_{\pm\pi/2}, Y_{\pm\pi/2}\}^{\otimes2} $ and apply an independent unitary error $U_{\epsilon} = \exp (-i \epsilon H_{\rm rand}  /2 )$ to each gate, where $H_{\rm rand}$ is a random Hamiltonian of unit norm.  The resulting error in the gate fidelity for each preparation or measurement pulse is approximatly $(\epsilon/2)^2$.  In table \ref{simtab} we simulate measurement data assusming a perfect identity gate.  Small errors in the analyzing gates can lead to significant errors in the tomgraphy.  In particular, errors of around $1 \%$ lead to the expected $3 - 4 \%$ errors in reconstruction, even if the gate being analyzed is a perfect identity transformation.  This error due to process tomography appears to become more severe as a ratio of the single gate error as the gates get more precise, and we conjecture that it will also scale poorly with the dimension of the system.